\documentclass[aip,reprint]{revtex4-2}
\pdfoutput=1

\usepackage{float}
\usepackage{color}
\usepackage{subfigure}
\usepackage{graphicx} 
\usepackage[T1]{fontenc}
\usepackage{amsmath,amssymb}
\usepackage{array}
\usepackage{color}
\usepackage{booktabs}
\usepackage[pdftex,bookmarks,colorlinks,breaklinks]{hyperref}  

\usepackage[textsize=scriptsize]{todonotes}

\begin{document}

%
%
\author{Martin P. Bircher}
\affiliation{Faculty of Physics
	\\University of Vienna
	\\Boltzmanngasse 5
	\\A-1090 Wien, Austria}
\author{Andreas Singraber}
\affiliation{Institute for Theoretical Physics
	\\TU Wien - Vienna University of Technology
	\\Wiedner Hauptstra\ss e 8-10
	\\A-1040 Wien, Austria}
\affiliation{Faculty of Physics
	\\University of Vienna
	\\Boltzmanngasse 5
	\\A-1090 Wien, Austria}
\author{Christoph Dellago}
\email{christoph.dellago@univie.ac.at}
\affiliation{Faculty of Physics
	\\University of Vienna
	\\Boltzmanngasse 5
	\\A-1090 Wien, Austria}

\title{Improved Description of Atomic Environments using Low-cost Polynomial Functions with Compact Support}

\begin{abstract}
The prediction of chemical properties using Machine Learning (ML) techniques calls for a set of appropriate
descriptors that accurately describe atomic and, on a larger scale, molecular environments. A mapping of
conformational information on a space spanned by atom-centred symmetry functions (SF) has become a standard technique
for energy and force predictions using high-dimensional neural network potentials (HDNNP). An appropriate choice
of SFs is particularly crucial for accurate force predictions. 
Established atom-centred SFs, however, are limited in their flexibility, since their functional form restricts
the angular domain that can be sampled without introducing problematic derivative discontinuities.
Here, we introduce a class of atom-centred symmetry functions based on polynomials with compact support called
polynomial symmetry functions (PSF), which enable a free choice of both, the angular and the radial domain covered.
We demonstrate that the accuracy of PSFs is either on par or considerably better than that of conventional,
atom-centred symmetry functions. 
In particular, a generic set of PSFs with an intuitive choice of the angular domain inspired
by organic chemistry considerably improves prediction accuracy for organic molecules in the gaseous and liquid phase,
with reductions in force prediction errors over a test set approaching 50\% for certain systems.
Contrary to established atom-centred SFs, computation of PSF does not involve any exponentials, and their intrinsic
compact support supersedes use of separate cutoff functions, facilitating the choice of their free parameters.
Most importantly,
the number of floating point operations required to compute polynomial SFs introduced here is considerably lower than that of other state-of-the-art
SFs, enabling their efficient implementation without the need of highly optimised code structures or caching, with
speedups with respect to other state-of-the-art SFs reaching a factor of 4.5 to 5.
This low-effort performance benefit substantially simplifies their use in new programs and emerging platforms
such as graphical processing units (GPU).
Overall, polynomial SFs with compact support
 improve accuracy of both, energy and force predictions with HDNNPs while enabling significant
speedups with respect to their well-established counterparts.
\end{abstract}

\maketitle

\section{Introduction}

In the past decade, computational chemistry has seen a tremendous increase in use of machine learning (ML) techniques
to overcome the length- and timescale problem burdening \emph{first principles} methods\cite{Dral2020JPCL,MaterCoote2019JCIM,ButlerDaviesEtAl2018N,Behler2016JCP}. As such, ML techniques
have seen particularly beneficial use in Molecular Dynamics (MD) simulations in the condensed phase.
By providing highly elaborate fitting mechanisms, ML allows for accurate interpolation between
training points on the potential energy surface (PES) generated from a computationally more expensive reference method.
This enables, \emph{e.g.}\ , MD simulations of water to be performed at the nano- rather than
the picosecond scale with forces and energies of \emph{first principles} accuracy\cite{MorawietzSingraberEtAl2016PNASUSA,ChengEngelEtAl2019PNASUSA}.
While still more expensive than classical force fields, ML techniques can 
incorporate information from the underlying PES well outside of local minima,
making the simulation of chemical reactions possible.
In contrast to reactive force-fields, no assumptions on the functional form of the interactions and therefore,
the underlying PES, have to be made\cite{SenftleHongEtAl2016NCM,BarcaroCarravettaEtAl2019APX}.
Instead, the fit is carried out mostly in a black-box manner,
with one of the most prominent adjustable input parameters being
a set of functions that uniquely describe the environment of every atom in a manner that 
is invariant both to permutation, rotation and translation\cite{Behler2016JCP,DeringerCaroEtAl2019AM,FaberHutchisonEtAl2017JCTC}.

Machine learning techniques commonly used to interpolate between \emph{first principles} data include kernel-based
methods\cite{BartokKondorEtAl2013PRB,BartokPayneEtAl2010PRL,RuppTkatchenkoEtAl2012PRL,ThompsonSwilerEtAl2015JCP} and neural network potentials (NNP)\cite{BlankBrownEtAl1995JCP,Behler2011PCCP}. 
A prominent framework of the latter is due to Behler and Parrinello\cite{BehlerParrinello2007PRL,Behler2015IJQC}, who advocated the use of high-dimensional
neural network potentials (HDNNP) in combination with atom-centred input functions.
In the context of HDNNPs, these functions are commonly referred to as symmetry functions (SF)\cite{Behler2011JCP}.
Ideally, these functions provide a unique description of the atomic environment of every structure in the training set.
Several functional forms have been proposed\cite{GeigerDellago2013JCP,GasteggerSchwiedrzikEtAl2018JCP}, which commonly include a radial and angular part multiplied by a cutoff function
that ensures that the former are naught outside of predefined bounds. SFs are therefore commonly a product of up to three
different types of functions, and parametrising them properly for a given chemical system can become non-trivial.
An intuitive construction of SFs can be based on structural features of the system at hand. For instance,
sufficient spatial resolution could be obtained by taking into account all relevant peaks in the radial distribution
function of the system and by covering all chemically relevant angles spanned by an atom and two of its neighbours.
The width and centre of a symmetry function should then reflect fluctuations of the quantity it describes in the training data.
Such a strategy has successfully been used, \emph{e.g.}\, for the construction of a neural
network of water and ice, including different ice phases\cite{MorawietzBehler2013JPCA,MorawietzBehler2013ZPC,MorawietzSingraberEtAl2016PNASUSA}.

However, whereas such a non-automated approach can yield symmetry function sets much smaller than those generated using
automated procedures\cite{GasteggerSchwiedrzikEtAl2018JCP,ImbalzanoAnelliEtAl2018JCP}, there are some pitfalls associated to an intuitive approach:
While the position of the maxima of radial functions can generally be freely chosen within the cutoff radius,
this is not necessarily the case for angular terms, which are commonly based on a cosine
and which are therefore restricted to peak at either $0$ or $\pi$. 
Introducing a simple phase shift in the cosine has been shown to work well for energy evaluations\cite{SmithIsayevEtAl2017CS},
but it would be bound to fail for force predictions, since any shift in a cosine will inevitably introduce
derivative discontinuities at $0$ and $\pi$:
The evaluation of angles is restricted to a domain of $[0,\pi]$, imposing symmetries in the angular functions
of the form $f(\vartheta) = f(-\vartheta)$ and $f(\pi+\vartheta) = f(\pi-\vartheta)$.
Phase shifts break this symmetry and introduce non-zero derivatives at $f(0)$ and $f(\pi)$. Since derivatives of the angular functions
are used to compute forces (\emph{vide infra}), this results in non-unique descriptions at $\vartheta = 0$ and $\vartheta = \pi$.

In order to account for
equilibrium angles that are not centred on either peak and which only fluctuate over a few degress -- such
as a double bond -- one therefore uses linear combinations of angular symmetry functions which are centred
on either $0$ or $\pi$, increasing the overall number of SFs used.
The presence of cutoff functions also obfuscates the choice of input SF parameters:
If the SF are constructed from stuctural properties such as the radial distribution function of a liquid,
the position of the radial peaks is not easily deduced from the input parameters, since the product of radial term and cutoff function
may exhibit a shifted maximum and a more rapid decay with respect to the Gaussian function alone. In particular, if the Gaussian
is not centred on the origin, this may lead to a symmetry function that is not symmetric with respect to its peak. Therefore,
it is usually necessary to plot every single symmetry function in order to monitor the correspondence between the input
parameters and the final system setup.
Hence, parametrising symmetry functions based on structural properties can be a cumbersome procedure.
While universal rules to generate input SFs have been proposed\cite{GasteggerSchwiedrzikEtAl2018JCP,ImbalzanoAnelliEtAl2018JCP}, 
they often increase the number of SF needed with respect
to an optimal, tailored set of parameters. This can hamper performance.

When using high-quality {first principles} methods to generate the training set, the {first principle} calculations
usually represent the computational bottleneck. However, both the training of the neural network itself as well as its use
in productive calculations carry a certain overhead that is typically larger than that of classical force fields.
The computational cost of the training is in part associated to the number
of symmetry functions used, as well as their computational complexity. While it is possible to store the values of all symmetry functions
during training\cite{SingraberMorawietzEtAl2019JCTC}, 
making their calculation necessary only once at the very beginning of the training procedure, this is evidently
impossible when the neural network potential is used to predict unknown structures \emph{e.g.}\ during a MD run.
Factors that contribute to the overhead of symmetry function evaluation are two-fold:
The most popular choice for the radial term are Gaussian functions. However, exponentials are comparably expensive to compute.
Furthermore, the calculation of a cutoff function (often cosines or hyperbolic tangents) introduces an additional overhead
with respect to bare radial functions. To this end, optimised cutoff functions based on polynomials have already been proposed.
Still, the disadvantages of the evaluation of exponentials for the radial terms, their possibly
becoming asymmetric due to multiplication with a cutoff function and
the impracticability of introducing phase shifts in the angular components remain. While strategies to reduce the large number
of floating point operations during symmetry function evaluation have been developed, \emph{e.g.} by grouping and
caching\cite{SingraberMorawietzEtAl2019JCTC,SingraberBehlerEtAl2019JCTC},
this involves intermediate storage of quantities, loop breaks and many \texttt{if}-statements, rendering their implementation
cumbersome, reducing code readability and making performance benefits both problem- and architecture dependent.

It would therefore be desirable to have a framework at hand where the number of floating point operations in SF evaluations 
is substantially reduced, where the SF remain symmetric with respect to their maxima,
where the input parameters directly reflect the shape of the resulting function, and where the maximum
of the angular component can be adjusted to reflect equilibrium angles. 
This would alleviate the need to plot every symmetry function during construction, making the resulting procedure more
effecient and less error-prone, greatly simplifying the choice of parameters based on structural
properties of the system at hand. Most importantly, a reduction of floating point operations supersedes the need
of complex code optimisation and renders the efficient implementation of symmetry functions architecture-independent.

In the following, we will demonstrate that further simplifications to existing symmetry functions
are possible by replacing the product-based ansatz of radial, angular and cutoff function by a
product of polynomials with compact support, making the use of cutoff functions obsolete.
Such polynomial symmetry functions have the advantage of allowing for angular terms to be centred anywhere within 
$\vartheta \in [0,\pi]$ without introducing derivative discontinuities at $\vartheta = 0$ and
$\vartheta = \pi$. The latter is of particular importance for force evaluations, as discontinutities in
 symmetry function derivatives will directly impact the forces computed.
 The evaluation of these polynomial symmetry functions does not involve exponentiation. Instead,
the most demanding low-level operation becomes the calculation of
an arccosine when computing angular terms. The use of one simple functional form for both angular and radial terms
considerably reduces the number of floating point operations associated to the evaluation of a single SF.
This does not only result in significant simplifications in the underlying
code, but comes at the benefit of speedups of up to a factor of about 4.5 to 5, while their performance
with respect to highly optimised cache and grouping based SF implementations still reaches a factor of almost 2 without
the need of further algorithmic optimisations.

This text is organised as follows: First, we briefly introduce the concept of HDNNPs
and describe the mathematical form of commonly used symmetry functions first proposed by Behler
and Parrinello\cite{BehlerParrinello2007PRL,Behler2016JCP}.
We then introduce our new polynomial symmetry functions (PSF).
We will go on to demonstrate that use of polynomial symmetry functions can improve accuracy of
neural network potentials when compared to conventional symmetry functions.
To this end, both dynamic and static properties will be compared for a water and copper sulfide model system.
By training a HDNNP for the rotation of the amino group of the organic dye
DMABN\cite{BircherRothlisberger2018JCTC,PeachBenfieldEtAl2008JCP}, we will show that the 
angular flexibility of polynomial symmetry functions simplifies the choice of angular parameters compared to Behler-Parrinello type
symmetry functions (BPSF). HDNNPs for liquid ethyl benzene and anisole that were trained using both PSF and PBSF
will further support this perspective. Finally, we will show that use of PSF in conjunction with recently
proposed weighted atom-centred symmetry functions (wACSF)\cite{GasteggerSchwiedrzikEtAl2018JCP} can further improve performance 
of HDNNPs for predicting enthalpies of formation of the QM9\cite{RamakrishnanEtAl2014Nature} database.
We will conclude the discussion by comparing execution times for the evaluation of the NNPs between BPSFs and PSFs,
showing that the execution time per symmetry function is substantially reduced for our polynomial ansatz.

\section{Symmetry Functions for High-Dimensional Neural Network Potentials}

\subsection{Neural Networks in a Nutshell}

First applications of neural network potentials in computational chemistry date back as far as 1995\cite{BlankBrownEtAl1995JCP}. 
A considerable advance in the prediction of molecular forces and energies was made in 2007, when Behler and Parrinello\cite{BehlerParrinello2007PRL} proposed
to use high-dimensional neural network potentials (HDNNP) in combination with a ficticious decomposition of the 
potential energy $V$ from $N$ individual atoms:
\begin{equation}
\label{eq:V}
V = \sum_i^N V_i(\mathbf{G}_i)
\end{equation}
The atomic contribution $V_i$ depends on the chemical environment of atom $i$, which is described by a tuple
of $N_j$ input values $\mathbf{G}_i = \{ G_1, \dots, G_j \}$. The forces on the $k$th atom follow from the gradient:
\begin{equation}
-\nabla_k V = - \sum_i^N \sum_j^{N_j} \frac{\partial V_i}{\partial G_{i,j}} \nabla_k G_{i,j}
\end{equation}

\begin{figure}
\input{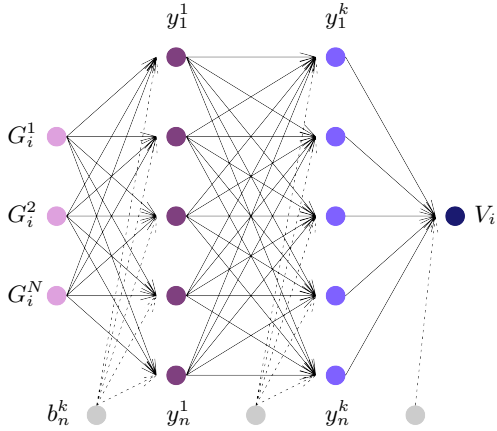}
\caption{Schematic representation of a Behler-Parrinello HDNNP for a given element $i$. Three symmetry functions
$G_i$ are used as input, and the atomic contribution to the total potential energy $V_i$ is predicted.
Labels correspond to Eq.\ \ref{eq:nnp}.}
\label{fig:nnp}
\end{figure}

The $\mathbf{G}_i$ are used as the input layer of a feed-forward neural network (Fig.\ \ref{fig:nnp})
that yields the scalar $V_i$. The network itself is constituted by multiple layers of nodes, which are connected
to each other by virtue of weights $a_{nm}$.
Starting from a set of input values $(\mathbf{G}_i)$, the data is propagated to node $n$ of hidden layer $k$ as follows:
\begin{equation}
\label{eq:nnp}
y_n^k = f_a \left( b_n^k + \sum_m^{M_l} a_{nm}^{lk} y_m^l \right),
\end{equation}
where $l=k-1$, $y_n^k$ represents the value of the node and $f_a$ is a non-linear activation function.
$b_n^k$ denotes the bias associated to $y_n^k$
and $M_l$ is the number of neurons in layer $l$.
One independent neural network per element is used. Together with the biases $b_n^k$, the weights $a_{nm}$ are the fitting
parameters of the HDNNP. 

The atomic environment descriptors $\mathbf{G}_i$ are commonly called symmetry functions (SF).
In order to ensure permutational symmetry, the symmetry functions are identical for all atoms of the same element.
However, in order to account for varying chemical environments (bond lengths, van-der-Waals radii, relevant bond angles),
they can differ between different elements. An appropriate choice of symmetry functions is crucial in order
to reliably discriminate between unique points on the PES. 

\begin{figure*}
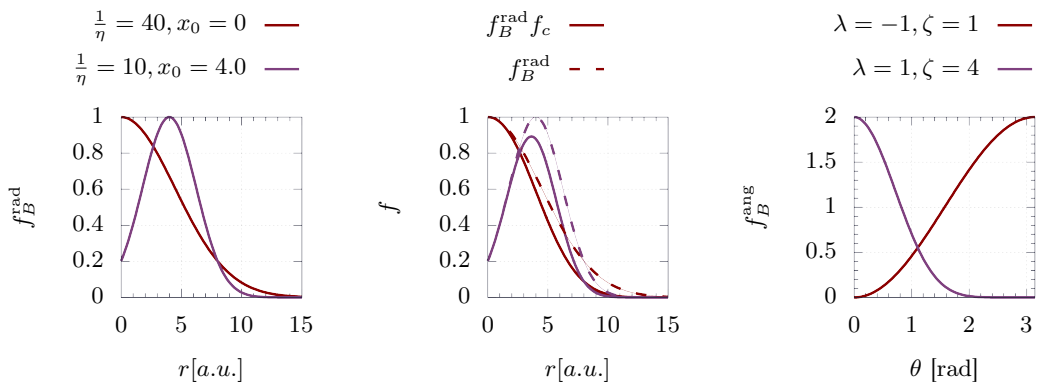

\input{fB_rad}
\input{fBfc_rad}
\input{fB_ang}
\caption{Radial (left), angular (middle) and cutoff (right) functions proposed by Behler. \emph{Left:} Due to the repulsive exchange wall,
         regions around $r=0$ are never sampled, and the radial function therefore need not be $0$ at $r = 0$. 
         \emph{Middle:} A simple cutoff function with continuous derivatives at $0$ and $r_c$ is multiplied with
                       the radial term to obtain a function with compact support. Note that this multiplication
                       shifts the maxima of Gaussians that are not centred on $r_0 = 0$.
         \emph{Right:} Minimum and maximum of the cosine are swapped for $\lambda = -1$. Specifying $\zeta > 1$ contracts
         the function towards its maximum.}
\label{fig:fB}
\end{figure*}

\subsection{Behler-Parrinello Symmetry Functions}
In the case of Behler-Parrinello HDNNPs, the input layer consists of scalars obtained from
a predetermined set of atom-centred symmetry functions.
The choice of symmetry functions should discriminate between all relevant structural features,
such that every distinct point on the potential energy surface is described by a unique tuple of
symmetry function values. In practice, this is achieved by combining a set of spherically symmetric and angle-dependent functions that
have to be parametrised appropriately. Some examples are given in Fig.\ \ref{fig:fB}.
For the radial part, Behler and Parrinello suggested:
\begin{equation}
f_\text{B}^\text{rad} (r,r_s) = e^{-\eta(r-r_s)^2},
\end{equation}
where $r$ is an interatomic distance and $r_s$ its reference point.
An angular dependency can by introduced by a function of the form:
\begin{equation}
f_\text{B}^\text{ang} (\vartheta,\lambda,\zeta) = \left(1 + \lambda \cos(\vartheta) \right)^\zeta,
\end{equation}
where $\vartheta$ is the angle formed by any three atoms, bound by $[0,\pi]$.
This function appropriately has
$\partial f_\text{B}^\text{ang} / \partial \vartheta \lvert_{\vartheta = 0} 
= \partial f_\text{B}^\text{ang} / \partial \vartheta \lvert_{\vartheta = \pi} = 0$. 
For $\lambda = 1$, maximum and minimum of the function lie at $0$ and $\pi$, respectively; for $\lambda = -1$, the inverse holds.
Chosing $\zeta > 1$ allows to contract the function, resulting in a more rapid decay. Values of $\zeta < 1$ are not possible,
since this would result in non-zero derivatives at the boundaries.

While the form of $f^\text{ang}_\text{B} (\vartheta,\lambda,\zeta)$ guarantees zero derivatives at the boundaries
for appropriate choices of $\lambda$ and $\zeta$, the exponential in $f^\text{rad}_\text{B} (r,r_c)$ is finite everywhere.
Therefore, a cutoff function with cutoff radius $r_c$ is introduced, which ensures that the radial term and its
derivatives are zero for all $r > r_c$. Common choices of cutoff function $f_c(r,r_c)$ include
\begin{align}
f_c(r,r_c) = \begin{cases}
\frac{1}{2} \left[ \cos\left(\frac{\pi r}{r_c}\right) + 1 \right] \quad &\text{for} \quad r \le r_c \\
0 \quad &\text{for} \quad r > r_c
\end{cases}
\end{align}
or
\begin{align}
f_c(r,r_c) = \begin{cases}
\tanh^3 \left(1 - \frac{r}{r_c}\right) \quad &\text{for} \quad r \le r_c \\
0 \quad &\text{for} \quad r > r_c
\end{cases}
\end{align}

\begin{figure*}
\input{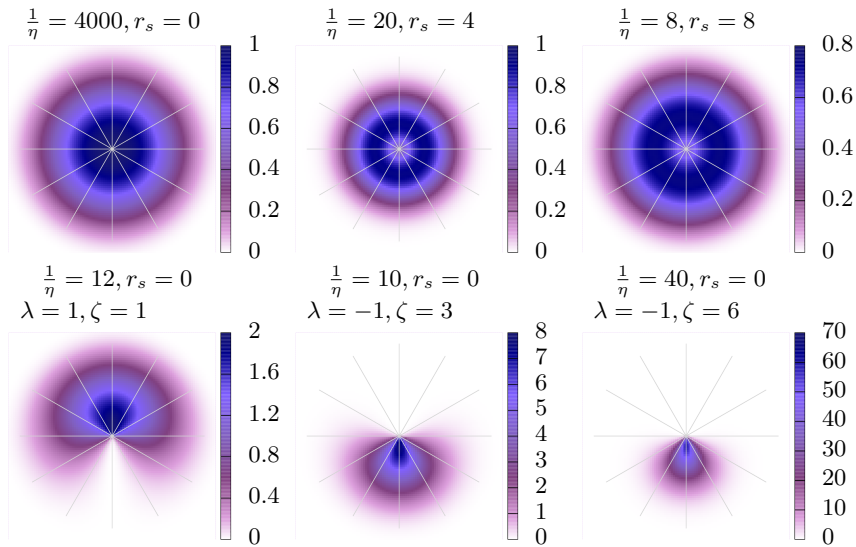}
\caption{Radial (upper rows) and angular (lower rows) symmetry functions of the Behler-Parrinello type. The reference atom $i$
         is positioned at the origin. For radial functions, the plot shows the dependency of $f^\text{rad}_B$ 
         on the position of neighbour $j$. For angular functions, it is assumed that one neighbouring atom $j$ has a fixed
         position at $x=0,y=1$, with the plot showing the dependency of $f^\text{ang}_B$ on the position
         of the second neighbour $k$.}
\label{fig:behler}
\end{figure*}

With this choice of functions, the radial symmetry function $G_i$ centred on an atom $i$ is given by 
a sum over all its neighbours $j$ that are within a $r_c$ of $i$:
\begin{equation}
G_i^\text{rad}(r_c,r_s) = \sum_{j \ne i} f^\text{rad}_{\text{B}} (r_{ij},r_s) f_c(r_{ij},r_c).
\end{equation}
Note that such a product may have a width and a maximum different from $f^\text{rad}_\text{B}$ alone,
see also Fig.\ \ref{fig:fB}.
The angular symmetry function is constructed from a product of angular and radial terms, with
the sum encompassing all neighbours $j$ and $k \ne j$.
Depending on the radial term, two types of angular symmetry functions can be distinguished,
namely narrow $n$,
\begin{widetext}
\begin{equation}
G_i^\text{ang.n} (r_c,r_s,\lambda,\zeta) = 2^{(1-\zeta)} \sum_{j \ne i, k > j}
     f^\text{rad}_{\text{B}} (r_{ij}, r_s) f^\text{rad}_{\text{B}} (r_{ik}, r_s) f^\text{rad}_{\text{B}} (r_{jk}, r_s)
     f^\text{ang}_{\text{B}} (\vartheta_{ijk},\lambda,\zeta)
     f_c(r_{ij},r_c) f_c(r_{ik},r_c) f_c(r_{jk},r_c),
\end{equation}
which also constrains the interatomic distance between atoms $j$ and $k$,
and the wide form $w$:
\begin{equation}
G_i^\text{ang.w} (r_c,r_s,\lambda,\zeta) = 2^{(1-\zeta)} \sum_{j \ne i, k > j}
     f^\text{rad}_{\text{B}} (r_{ij}, r_s) f^\text{rad}_{\text{B}} (r_{ik}, r_s) 
     f^\text{ang}_{\text{B}} (\vartheta_{ijk},\lambda,\zeta)
     f_c(r_{ij},r_c) f_c(r_{ik},r_c),
\end{equation}
\end{widetext}
the evaluation of which is more computationally expedient due to the lack of a radial term for $jk$.
These symmetry functions have successfully been applied in a wide variety of contexts.
Note that the distribution of symmetry function values at $\lambda = -1$ is not symmetric
with respect to $\lambda = 1$ due the presence of a radial function\cite{GasteggerSchwiedrzikEtAl2018JCP}.
Fig.\ \ref{fig:behler} displays the dependency of BPSFs on the position of atomic neighbour(s).

Recently, Singraber \emph{et al.}\ have introduced a simple expression for $f_c$ based on polynomials\cite{SingraberBehlerEtAl2019JCTC}:
\begin{align}
\label{eq:fcpoly}
f_c^\text{poly} (r,r_c) = \begin{cases}
f^\text{poly2} \left( \frac{r}{r_c} \right) \quad &\text{for} \quad 0 \le r \le r_c \\
0 \quad &\text{for} \quad r > r_c
\end{cases}
\end{align}
with the polynomial:
\begin{equation}
\label{eq:fpoly}
f^\text{poly2} (x) = x^3 ( x (15 - 6x) - 10 ) + 1
\end{equation}
with $f^\text{poly2}(0) = 1$ and $f^\text{poly2} (1) = 0$. 
At $x=0$ and $x=1$, Eq.\ \ref{eq:fpoly} has continuous derivatives up to second order, which ensures
a continuous, smooth transition to outside of the compact set.
Similar polynomials with continuous derivatives up to even higher order can be easily constructed\cite{SingraberBehlerEtAl2019JCTC},
but their use does not result in any practical benefit\cite{Behler2015IJQC}.
The main advantage of polynomial cutoff functions as
introduced in Ref.\ \citenum{SingraberBehlerEtAl2019JCTC}
lies in their expedient evaluation compared to expressions based on hyperbolic tangent
or cosine, reducing the overhead due to the cutoff function.
However, the computation of costly exponentials is still needed for all radial components.
Hence, a grouping strategy was proposed in Ref.\ \citenum{SingraberBehlerEtAl2019JCTC} which
avoids multiple evaluations of the same exponential term, resulting in considerable speedups.

\subsection{Polynomial Symmetry Functions with Compact Support}

\begin{figure*}
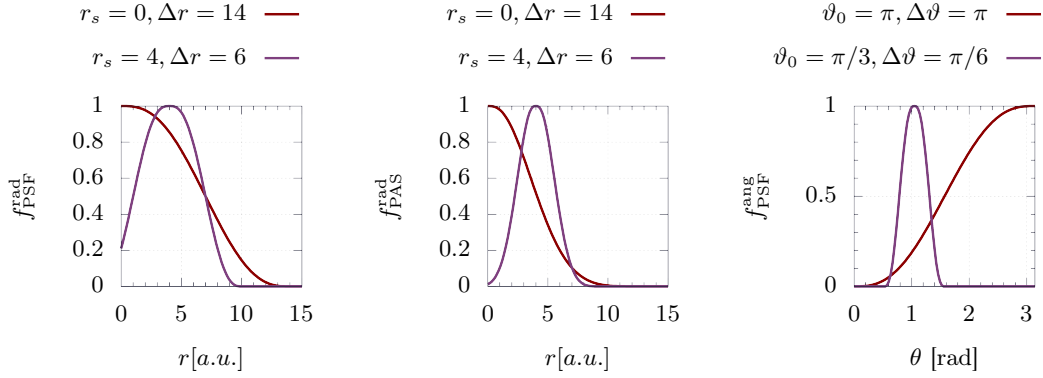

\input{PSF_rad}
\input{PAS_rad}
\input{PSF_ang}
\caption{Radial PSF (left), radial PAS (middle) and angular (right) SFs with compact support proposed here.
         \emph{Left:} Radial function with point-symmetry around $0.5 r_\text{max}$. Note that no cutoff
                      function is needed, which enables a precise choice of maxima and shifts.
         \emph{Middle:} Asymmetryic radial function with longer tail, with a shape comparable to the product of a Gaussian
                       and a cutoff function in Fig.\ \ref{fig:fB}.
         \emph{Right:} Angular polynomial SF. Note that the angular domain can be freely chosen and does not need to 
                       have maxima at $0$ or $\pi$.}
\label{fig:PSF}
\end{figure*}

On one hand, the summation over the arguments of $G^\text{ang}$ still involves
a substantial number of exponential functions that have to be explicitly evaluated even when a grouping strategy is used.
On the other hand, the present choice of $f^\text{ang}$ implies that, in order to specifically sample
angles which are not centred close to either $0$ or $\pi$, a combination 
of several values of $\lambda$ and $\zeta$ is necessary, since introducing an angular shift
$\vartheta_0$ would inevitably violate the boundary condition $\partial f^\text{ang} / \partial \vartheta = 0$
at the bounds of the angular domain. In the following, we will attempt to overcome this issue
by constructing a simple set of symmetry functions based on polynomials with compact support which
can be used to indiscriminately describe both anglular and radial dependencies.

In order to do so, we generalise Eq.\
\ref{eq:fcpoly} to a generic form, expressed in terms of the boundaries $x_\text{min},x_\text{max}$ of the
underlying domain. In particular, for our polynomial symmetry functions $f_\text{PSF}$, we also allow for arguments $x < 0$:
\begin{widetext}
\begin{align}
\label{eq:PSF}
f_\text{PSF} (x,x_0,\Delta x) = \begin{cases}
f^\text{poly2} \left( \frac{x-x_0}{\Delta x} \right) \quad &\text{for} \quad x_0  \le x \le x_0 + \Delta x \\
f^\text{poly2} \left( \frac{x_0-x}{\Delta x} \right) \quad &\text{for} \quad x_0 - \Delta x  \le x < x_0 \\
0 \quad &\text{elsewhere}
\end{cases}
\end{align}
where $x_0 = \frac{1}{2}(x_\text{max}+x_\text{min})$ and $\Delta x = \frac{1}{2}(x_\text{max}-x_\text{min})$.
The resulting function is symmetric with respect to $x_0$ and has continuous derivatives -- up to an order determined
by the construction of $f^\text{poly2}$ -- on all $\mathbb{R}^3$.
It is possible to further influence the decay of the radial function by appropriately modifying
its argument such that the symmetry function decays more rapidly around its maximum while also approaching $0$
more slowly, thus breaking the point symmetry around $f_\text{PSF}(0.5)$. We therefore define the asymmetric polynomial symmetry function (PAS) as:
\begin{align}
\label{eq:aPSF}
f_\text{PAS} (x,x_0,\Delta x) = \begin{cases}
f^\text{poly2} \left( \frac{2(x-x_0)}{\Delta x}  - \left( \frac{x-x_0}{\Delta x} \right)^2 \right)
   \quad &\text{for} \quad x_0  \le x \le x_0 + \Delta x \\
f^\text{poly2} \left( \frac{2(x_0-x)}{\Delta x}  - \left( \frac{x_0-x}{\Delta x} \right)^2 \right)
   \quad &\text{for} \quad x_0 - \Delta x  \le x < x_0 \\
0 \quad &\text{elsewhere}
\end{cases}
\end{align}
which offers the same benefits as $f_\text{PSF}$ for $x \in [0,1]$.
By construction, $f_\text{PAS}$ resembles the shape of the product of a local, narrow Gaussian function with a
cutoff function and can therefore be used to adapt existing BPSF-setups to polynomial form.
The functional forms of $f_\text{PAS}$ and $f_\text{PSF}$ are depicted in Fig.\ \ref{fig:PSF}.

\begin{figure*}
\input{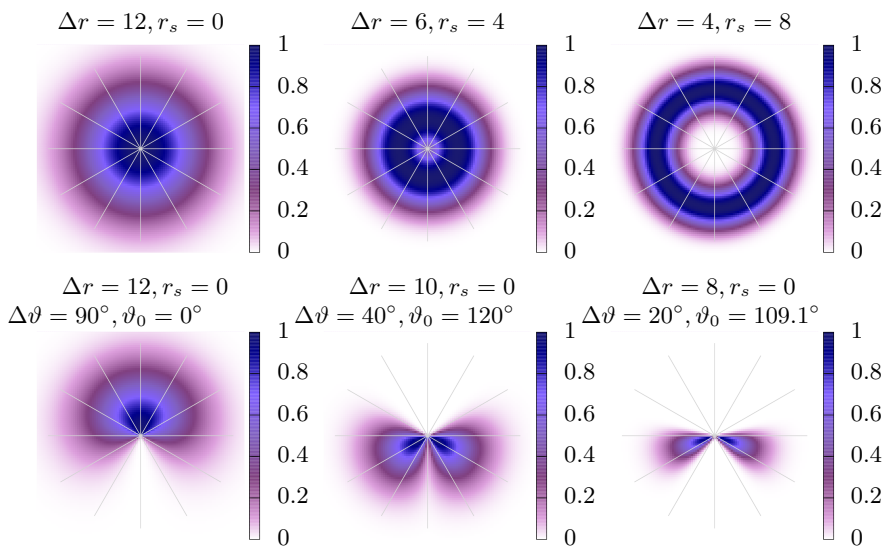}
\caption{Radial (upper rows) and angular (lower rows) symmetry functions of the PAS type. The reference atom $i$
         is positioned at the origin. For radial functions, the plot shows the dependency of $f^\text{rad}_B$ 
         on the position of neighbour $j$. For angular functions, it is assumed that one neighbouring atom $j$ has a fixed
         position at $x=0,y=1$, with the plot showing the dependency of $f^\text{ang}_B$ on the position
         of the second neighbour $k$. Note that with PSF and PAS it becomes possible to cover specific angular
         ranges associated, \emph{e.g.} to double bonds ($60^\circ$ angle, lower middle) or tetrahedra 
         ($109.1^\circ$, lower right).}
\label{fig:poly}
\end{figure*}

The corresponding radial symmetry functions can now be constructed as:
\begin{equation}
\label{eq:Grp}
G_i^\text{rad/P} = \sum_{j \ne i} f_\text{P} (r_{ij},r_s,\Delta r).
\end{equation}
with $f_\text{P}$ either $f_\text{PSF}$ or $f_\text{PAS}$.
Their narrow angular counterparts read:
\begin{equation}
\label{eq:Grangn}
G_i^\text{ang.n/P} = \sum_{j \ne i, k>j}
      f_\text{P}   (r_{ij},r_s,\Delta r)
      f_\text{P}   (r_{ik},r_s,\Delta r)
      f_\text{P}   (r_{jk},r_s,\Delta r)
      f_\text{PSF} (\vartheta_{ijk},\vartheta_0,\Delta \vartheta)
\end{equation}
while the wide angular functions become:
\begin{equation}
\label{eq:Grangw}
G_i^\text{ang.w/P} = \sum_{j \ne i, k>j}
      f_\text{P}   (r_{ij},r_s,\Delta r)
      f_\text{P}   (r_{ik},r_s,\Delta r)
      f_\text{PSF} (\vartheta_{ijk},\vartheta_0,\Delta \vartheta)
\end{equation}
\end{widetext}
where the radial component $f_\text{P}$ can again be described by both $f_\text{PAF}$ or $f_\text{PSF}$.
Both types of angular symmetry function can be centred anywhere within $[0,\pi]$, provided
that $\partial f_\text{PSF} / \partial \delta \lvert_{\vartheta=0} = 0$ and
$\partial f_\text{PSF}/\partial \delta \lvert_{\vartheta=\pi} = 0$.
This enables chemically intuitive choices of angular maxima, \emph{i.e.}\ $60^\circ$ for
a double bond or $109,4^\circ$ for a carbon-carbon single bond.
Fig.\ \ref{fig:poly} shows the dependency of PAS-type symmetry functions on the position of atomic neighbour(s).

Eqs \ref{eq:PSF} to \ref{eq:aPSF} do not involve any exponentials, and the most expensive
arithmetic operation, an inverse cosine, has a modest computational footprint.

\begin{figure*}
\centering
\includegraphics[width=1.0\linewidth]{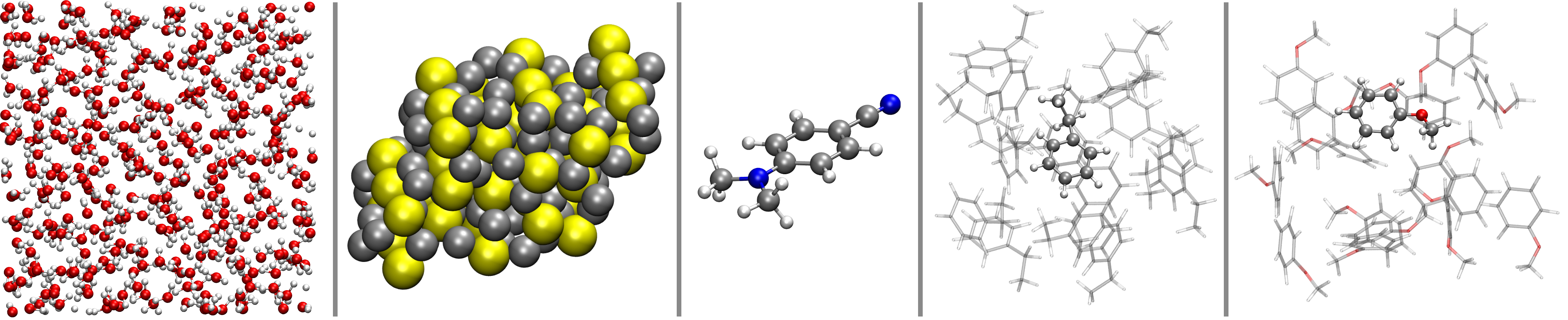}
\caption{Systems studied here. From left to right: Water, Cu$_2$S, DMABN, liquid ethyl benzene and anisole.}
\label{fig:overview}
\end{figure*}

\section{Computational Methods}
\subsection{Training of the HDNNP}
All HDNNPs were trained using the n2p2 package\cite{SingraberBehlerEtAl2019JCTC,SingraberMorawietzEtAl2019JCTC,n2p2}.
HDNNPs for the prediction of enthalpies of formation in the QM9\cite{RamakrishnanEtAl2014Nature} database were
obtained using the protocol of Ref. \citenum{GasteggerSchwiedrzikEtAl2018JCP} with five-fold cross validation, but
without dataset normalisation.
For the remaining setups,
10 training runs per system are carried out with an independent seed for the random number generator each.
If the value $G^i$ of a given symmetry function never exceeds $10^{-3}$ over the whole test set, the function is discarded before
training. The choice of $G^i$ is detailed in the Supporting Information.
Results discussed in this text refer to the best set of weights out of 10
independent runs (5 for the QM9 set), 
which are defined as the weights that yield the lowest force root mean-square deviation (RMSD) with respect
to a test set constituted by 10\% of training-set structures that were randomly removed before training. If several sets of weights yield
force RMSDs that are identical to a few \%, the set that displays the lowest relative error in energy predictions is used.
\subsection{Training Sets}
The systems investigated here are depicted in Fig. \ref{fig:overview}.
The training sets for water and copper sulphide correspond to those published in Refs \citenum{MorawietzSingraberEtAl2016PNASUSA}
and \citenum{SingraberMorawietzEtAl2019JCTC}, respectively.
Forces and energies for an isolated DMABN molecule, liquid ethyl benzene and anisole (16 molecules in a periodic box)
were obtained from high-temperature
Car-Parrinello Molecular Dynamics runs performed with the CPMD code\cite{CPMD}
using the SCAN exchange-correlation functional\cite{SunRuzsinszkyEtAl2015PRL}.
The amino group of DMABN
was rotated using a slowly growing harmonic restraint, giving rise to a set of non-equilibrium conformers.
Details on the computational setup can be found in the Supporting Information.

\begin{figure*}
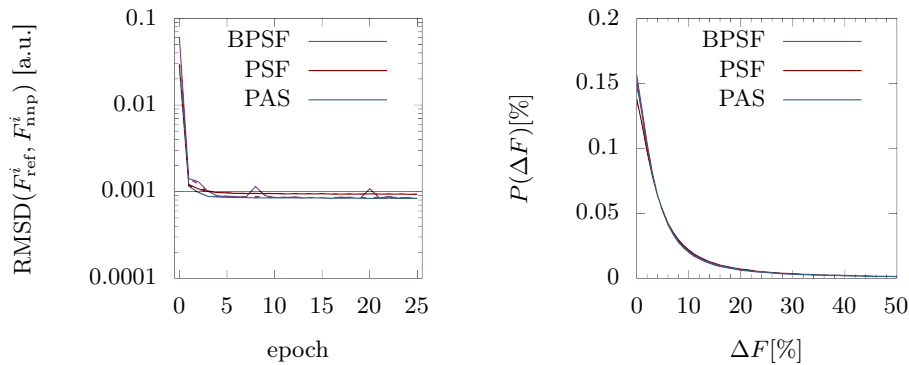

\hrule
\input{wat_F}
\input{wat_zoom}
\caption{Force predictions on water set of Ref.\ \citenum{SingraberMorawietzEtAl2019JCTC}. \emph{Ref} denotes references forces and \emph{nnp} the prediction
         by the HDNNP.
         \emph{Left:} Logarithmic learning curve showing the decrease of the RMSD between predicted and reference forces
                      as a function of training steps (epochs). There are no appreciable differences between
                      BPSF, PSF and PAS. A thin line denotes the target RMSD of 50 meV / \AA.
         \emph{Right:} Distribution $P(\Delta F)$ of relative force errors, $( \Delta F = F_\text{nnp} - F_\text{ref} ) / F_\text{ref}$
                       for BPSF, PSF and PAS.}
\label{fig:wat_train}
\end{figure*}

\subsection{Test Sets}
Test sets consists of 10\% randomly chosen configurations that are removed from the training set
before training and which are then used to validate the predictive power of the neural network potential.

\subsection{Molecular Dynamics using Neural Network Potentials}
All MD simulations using HDNNPs were carried out using the n2p2-interface\cite{SingraberBehlerEtAl2019JCTC,n2p2} to the LAMMPS code\cite{Plimpton1995JCP,LAMMPS}.
System setups are detailed in the Supporting Information.

\section{Results and Discussion}
\subsection{Water}

We first assess the accuracy that can be attained with polynomial SFs by comparing the efficiency and
accuracy of both training and productive simulations using identically set up HDNNPs.
Two setups of polynomial SFs will be used: One, PSF, consisting entirely of symmetric functions of the form
\ref{eq:PSF}, and one mixed setup, PAS, including both symmetric and asymmetric radial components.
We will compare against a reference setup using BPSFs,
which has successfully been used in Ref.\ \citenum{SingraberMorawietzEtAl2019JCTC} to describe the PES of water and several ice phases.
PSF and PAS were constructed by plotting the BPSFs and using their peaks and the full
width at half maximum as rough indicators for the polynomial input. For sake of simplicity, the only symmetric
radial terms (PSF-type) in the PAS set were those with maximum width; all others were chosen to be of the antisymmetric type.

\begin{figure*}
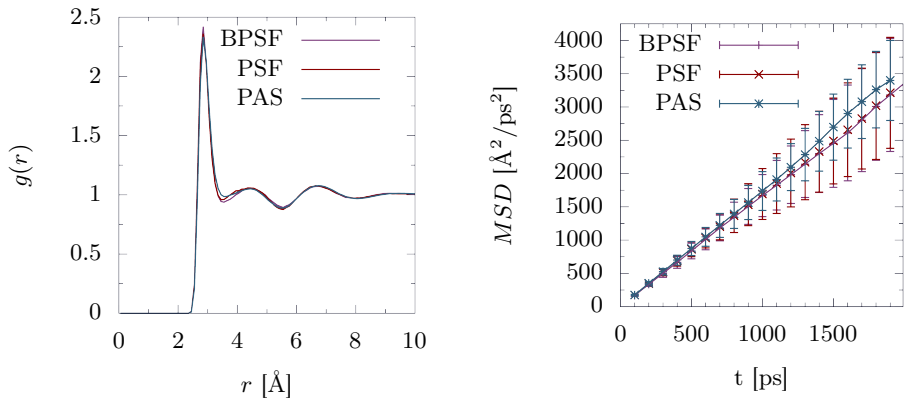

\input{gr}
\input{msd}
\caption{Properties of water obtained during a molecular dynamics run with HDNNPs trained using BPSF, PSF and PAS.
         \emph{Left:} Oxygen-oxygen radial distribution function $g(r)$ averaged over a 2 ns NVT trajectory. $g(r)$ obtained using
                      different symmetry functions practically show no difference.
         \emph{Right:} Mean-squared displacements and errors obtained over 2 ns of MD in the NVE ensemble. Error margins for all
                       lines overlap, suggesting equivalency between all symmetry functions investigated here.
         }
\label{fig:wat_perf}
\end{figure*}

First, we compare the learning curves of both methods to assess the efficiency of the training procedure. Here, we
define the learning curve as the RMSD between trained energy and forces and their reference
as a function of the number of training steps (epochs).
Curves for Behler-Parrinello type and polynomial SF setups are plotted in Fig.\ \ref{fig:wat_train}
and are virtually indistinguishable, suggesting that both types of symmetry
functions can be trained with equal efficiency.
RMSDs at the last training step are comparable between different methods, with the largest remaining force deviation
being observed for the PSF setup. Conversely, the lowest force RMSD is observed for the PAS setup.
BPSFs yield a training error intermediate between PAS and PSF data.

\begin{figure}[b!]
\input{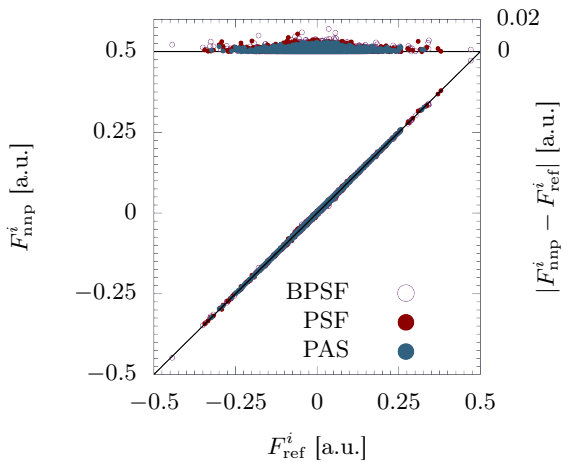}
\caption{Predicted (nnp) \emph{vs.}\ reference (ref) forces as obtained using BPSF, PSF and PAS.
         A deviation from the diagonal indicates an error in the force prediction. PAS and PSF further improve 
         the description of the forces largest in magnitude. Absolute deviations from the diagonal are plotted in the top panel.}
\label{fig:wat_diag}
\end{figure}

The same conclusion holds when analysing the distribution of both force and energy errors within the test set:
Force RMSDs are 43.7 meV/\AA{} for BPSFs, 48.5 meV/\AA{} for PSF and 43.2 meV/\AA{} for PAS, respectively.
Energy RMSDs range from 1.17 meV fo BPSFs to 0.95 meV for PAS.
In practical applications, these values could be considered identical.
In Fig.\ \ref{fig:wat_train}, we histogram both the relative force error for HDNNPs trained with
Behler-Parrinello type and polynomial SFs.
The resulting distributions for the test set compare well between PSF, PAS and BPSF: 
All relative errors show a rapid decay, indicating an excellent quality of the fit. PAS and BPSF
show very similar relative error distributions, whereas the distribution of PSF skews very lightly towards the right; however,
we will show that this is of no practical relevance.
Fig.\ \ref{fig:wat_diag} shows predicted forces against their reference values;
ideally, all values should lie on a diagonal. Use of PAS and PSF improves the accuracy of the most extreme force
values at either end of the graph, which lie closer to the diagonal than for BPSF.
This indicates that, although force RMSDs differ only little between the different SF setups, qualitative
improvements to conventional BPSFs are possible by use of a PSF or PAS set. 

Fig.\ \ref{fig:wat_perf} provides a practical comparison of the predictive power of the different networks.
Both the radial distribution function (left panel) as well as the mean square displacement as a function of time,
from which the diffusion coefficient can be derived,
(right panel) are practically indistinguishable between BPSF, PAS and PSF.
These results further support that polynomial SFs are a viable alternative to the commonly used Behler-Parrinello type functions.
The excellent agreement between
radial distribution functions and mean-squared displacements obtained from Molecular Dynamics suggest that the small differences
in force RMSD over the test set do not entail significant changes in the quality of the predictions made by the HDNNP.
Notably, the mean-squared displacements (MSD) for PSF and BPSF
are virtually identical.

\begin{figure*}
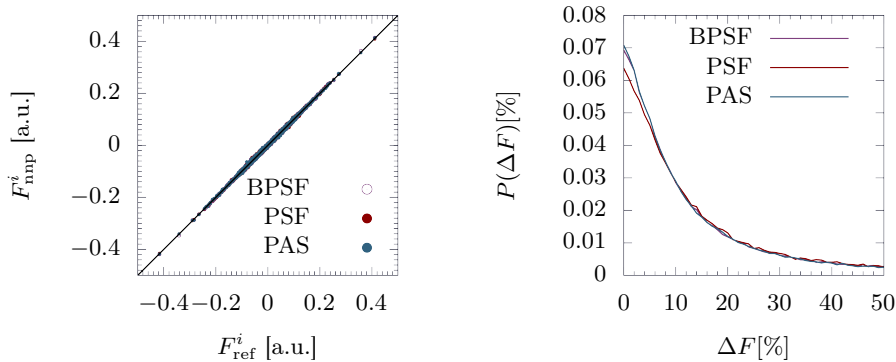

\centering
\input{cu2s_diag}
\input{cu2s_zoom}
\caption{Force predictions on the Cu$_2$S test set corresponding to 10\% of the structures of the training set
         of Ref.\ \citenum{SingraberMorawietzEtAl2019JCTC}. \emph{Ref} denotes references forces and \emph{nnp} the prediction
         by the HDNNP.
         \emph{Left:} Predicted (nnp) \emph{vs.}\ reference (ref) forces as obtained using BPSF, PSF and PAS.
         Shown are the forces largest in magnitude.
         A deviation from the diagonal indicates an error in the force prediction. PAS and PSF are better able to train the largest
         forces in the test set, which is reflected in a smaller deviation from the diagonal.
         \emph{Right:} Distribution $P(\Delta F)$ of relative force errors, $( \Delta F = F_\text{nnp} - F_\text{ref} ) / F_\text{ref}$
                       for BPSF, PSF and PAS.}
\label{fig:cu2s_train}
\end{figure*}

\begin{figure*}
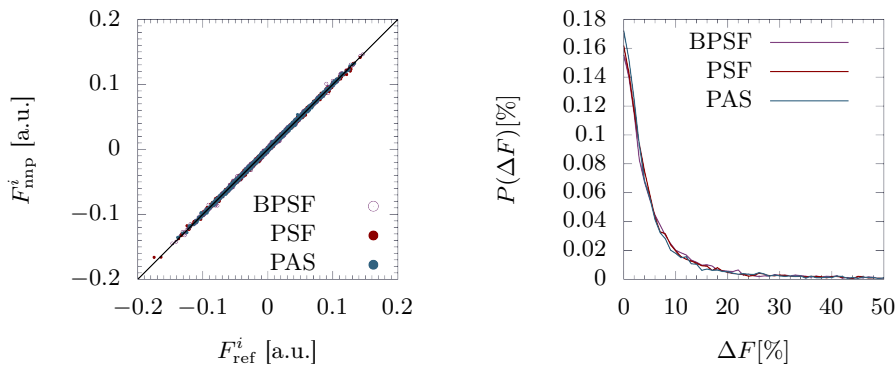

\centering
\input{dmabn_diag}
\input{dmabn_zoom}
\caption{Force prediction quality for an isolated DMABN molecule, including a rotation of its amino group along a non-equilibrium path.
         \emph{Left:} Predicted (nnp) \emph{vs.}\ reference (ref) forces as obtained using BPSF, PSF and PAS.
         A deviation from the diagonal indicates an error in the force prediction. PAS forces remain closer to the diagonal than either
         BPSF and PSF.
         \emph{Right:} Distribution $P(\Delta F)$ of relative force errors, $( \Delta F = F_\text{nnp} - F_\text{ref} ) / F_\text{ref}$
                       for BPSF, PSF and PAS. Whereas the distribution for BPSF and PAS are similar,
                       PSF again skew slightly towards the right.}
\label{fig:dmabn_train}
\end{figure*}

\subsection{Copper sulfide}
All symmetry function setups used for water were based on the optimised, hand-selected set of Ref.\ \citenum{MorawietzSingraberEtAl2016PNASUSA}.
Such a selection by hand may not always be possible. In the following, we shall further validate the accuracy of PSF and PAS
by comparing their performance to the accuracy of a semi-automatically generated set of BPSFs for copper sulfide.
It has been show in Ref.\ \citenum{SingraberMorawietzEtAl2019JCTC} that a HDNNP with BPSFs can be successfully used to observe 
structural phase transitions in this compound.

Fig.\ \ref{fig:cu2s_train} compares absolute and relative force errors obtained using BPSFs of Ref.\ \citenum{SingraberMorawietzEtAl2019JCTC}
as well as our PSF and PAS. Force RMSDs are 51.9 meV/\AA{} for BPSF, 56.7 meV/\AA{} for PSF and 52.3 meV/\AA{} for PAS,
which corresponds roughly to a 10\% difference between PSF and BPSF. 
Again, the performance of BPSF and PAS is on par, whereas use of PSF leads to slightly larger
errors, which is also reflected in the relative error distribution skewing lightly to the right. 
Overall, however, this effect
is expected to remain negligible in practical applications. Deviations from the diagonal
in the graph comparing prediction and reference forces (Fig.\ \ref{fig:cu2s_train}, right panel)
do not exhibit visible differences, further indicating that the general
performance of all symmetry function types investigated here remains comparable and that the differences in force RMSDs observed
here are not expected to influence the quality of the predictions in a significant manner.
Energy RMSDs remain similar for all methods, spanning
a range from 0.99 meV for PAS to 1.06 meV for BPSF.

\subsection{Organic molecules: Gas phase and liquids}
\label{sec:organic}
In the following, we shall compare the performance of BPSFs generated according to Ref.\ \citenum{ImbalzanoAnelliEtAl2018JCP}
with our polynomial symmetry functions using an isolated (gas-phase) DMABN molecule and liquid ethyl benzene.
We note that optimal symmetry function setups could possibly
be found for all SF types investigated here; however, since this is a time-intensive procedure,
 the main purpose of this comparison is to compare the out-of-the-box performance of some generic sets of SFs.

\begin{figure*}
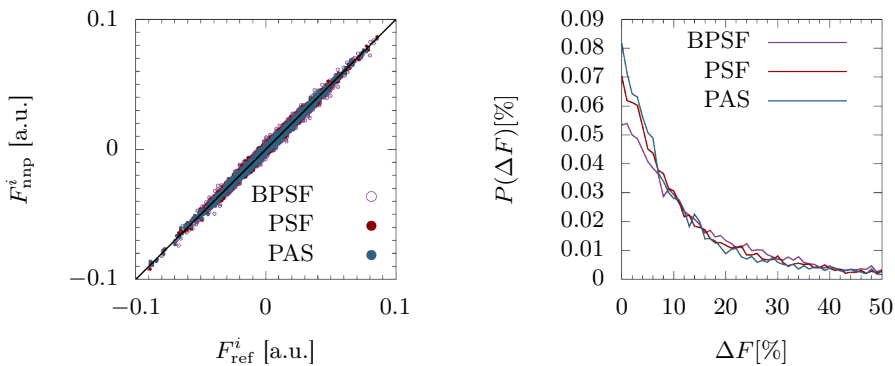

\centering
\input{etph_diag}
\input{etph_zoom}
\caption{Force prediction for ethyl benzene in its liquid state.
         \emph{Left:} Predicted (nnp) \emph{vs.}\ reference (ref) forces as obtained using BPSF, PSF and PAS.
         A deviation from the diagonal indicates an error in the force prediction. PAS forces remain closer to the diagonal than either
         BPSF and PSF.
         \emph{Right:} Distribution $P(\Delta F)$ of relative force errors, $( \Delta F = F_\text{nnp} - F_\text{ref} ) / F_\text{ref}$
                       for BPSF, PSF and PAS. Whereas the distribution for PSF and PAS are similar,
                       BPSFs skew considerably to the right, indicating a substantial difference in accuracy.}
\label{fig:etph_train}
\end{figure*}

\begin{figure*}
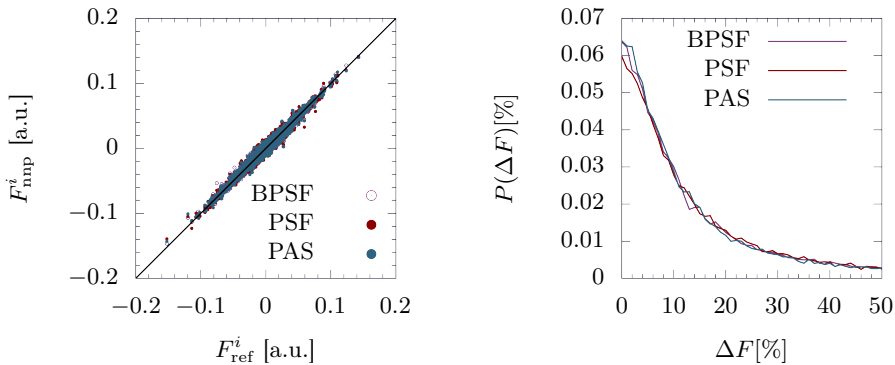

\centering
\input{anisole_diag}
\input{anisole_zoom}
\caption{Force prediction for liquid anisole.
         \emph{Left:} Predicted (nnp) \emph{vs.}\ reference (ref) forces as obtained using BPSF, PSF and PAS.
         A deviation from the diagonal indicates an error in the force prediction. PAS forces remain closer to the diagonal than either
         BPSF and PSF.
         \emph{Right:} Distribution $P(\Delta F)$ of relative force errors, $( \Delta F = F_\text{nnp} - F_\text{ref} ) / F_\text{ref}$
                       for BPSF, PSF and PAS. Whereas the distribution for BPSF and PAS are similar,
                       PSF skew slightly towards the right. However, this difference is much less pronounced than the one observed
                       for BPSF in Fig.\ \ref{fig:etph_train}.}
\label{fig:anisole_train}
\end{figure*}

To this end, we introduce a simple generation scheme for radial symmetry function parameters given a minimal
and maximum width, $\Delta_\text{min}$ and $\Delta_\text{max}$,
and a maximum cutoff radius $r_c$. For $N_0$ symmetry functions which are symmetric w.r.t
to the origin, we choose the width $\Delta^i_0$ parameters of the $i$-th out of $N_0$ PSF or PAS as:
\begin{align}
\Delta^0_0 &= r_c \\
\Delta^i_0 &= \Delta_\text{max} - \frac{(i-1)(\Delta_\text{max} - \Delta_\text{min})}{N_0-2} \quad ; \quad i \ge 1
\end{align}
For a given number $N_s$ of symmetry functions which are centred at $r_s^i \ne 0$, we obtain shifts and widths $\Delta^i_s$
as:
\begin{align}
\Delta^i_s &= \frac{2r_c}{N_s+1} \\
r_s^i &= \frac{(i-1)\Delta^i_s}{2}
\end{align}
This partitioning is reminiscent to the one introduced by Gastegger \emph{et al} in Ref. \citenum{GasteggerSchwiedrzikEtAl2018JCP}. 
Angles were chosen to reflect common structural features
encountered in organic chemistry: Linearity for triple bonds, $60^\circ$ and $109.1^\circ$ angles to describe double- and single carbon-carbon
bonds, respectively. This resulted in the following angular functions: $\vartheta_0^{1} = 0^\circ$, $\vartheta_1^{1} = 90^\circ$
and $\vartheta_3^{1} = 180^\circ$ with $\Delta \vartheta^{1,2,3} = 90^\circ$; $\vartheta_0^{4} = 60^\circ$, $\vartheta_0^{5} = 120^\circ$
and $\vartheta_0^6 = 109.1^\circ$ with $\Delta \vartheta^{4,5,6} = 60^\circ$ each.
For sake of comparison, the cutoff function distance for BPSFs and the maximum extent
of PAS and PSF were set to 15 a.u.

Structures in the DMABN-set cover regions of the PES thermally accessible at 1320K, including a forced rotation around the amino group
along a non-equilibrium path. This results in many forces with large magnitude which are well outside of the equilibrium regime in which
simple harmonic potentials can also yield good approximations. The DMABN set therefore serves as an assessment for the transferability
of our polynomial SFs to structures which are considerably off-equilibrium.

Force RMSDs on the test set amount to 63.2 meV/\AA{} for BPSF, 56.6 meV/\AA{} for PSF and 54.0 meV/\AA{} for PAS.
This difference is directly reflected in the right panel of Fig.\ \ref{fig:dmabn_train}, where the force histograms show an increasing
spread towards larger values when going from PAS over PSF to BPSF. Note that this amounts to a difference exceeding 15\% between
worst (BPSF) and best (PAS) set. Energy RMSDs range from 0.64 meV for PSF to 0.72 meV for BPSF, with the overall error remaining low for all types of functions.
The force predictions obtained here imply that for a generic test set containing a significant amount of off-equilibrium
structures, generic PAS and PSF sets are able to outperform BPSFs.

We will now show that the same considerations also hold for an organic liquid in which van der Waals interactions dominate.
For a test set of liquid ethyl benzene, using the same symmetry function setups as for DMABN,
we find force prediction RMSDs of 89.5 meV/\AA{} for BPSFs,
68.9 meV/\AA{} for PSF and 61.7 meV/\AA{} for PAS, respectively. This corresponds to a difference 
in accuracy approaching 50\% between BPSF and PAS.
Whereas these considerable differences are not evident from the diagonal force-to-force
plot in the left panel of Fig. \ref{fig:etph_train}, they become visible in the force histogram in the right panel: 
Histograms for PSF and PAS are very similar, with the only significant difference being the peak around 0\%, but BPSFs
 skew considerably to the right, and much more so than what was observed for PSF in the case of water or Cu$_2$S.
Energy RMSDs are about 0.05 meV for all SFs studied here, which is of excellent accuracy.
For liquid anisole, all symmetry function setups result in about equally accurate predictions: Observed force prediction
RMSDs amount to 100.8 meV/\AA{} for BPSFs and a comparable 106.5 meV/\AA{} for PSF,
with the lowest deviation again being observed at 98.7 meV/\AA{} for PAS. All SFs yield energy RMSDs that can be
considered identical, within a small range from 0.08 meV for BPSF to 0.11 meV for PAS and PSF.
Overall, the trends obtained for force predictions in organic molecules indicate
that our generic PAS and PSF setups on average exhibit superior transferability between systems, while at the same
time retaining lower force errors on the test sets compared to generic BPSFs. Importantly, whereas BPSFs
performed considerably worse than polynomial functions for liquid ethyl benzene, neither PAS nor PSF showed a considerably
inferior performance with respect to BPSF in any of the other systems investigated here.
Energy predictions are much more robust with respect to the choice of system, they remain of comparable
accuracy for all SFs investigated here. 

\subsection{QM9 Database}
In order to pinpoint possible improvements when solely energies are to be treated, we have constructed a set of
weighted atom-centred symmetry functions (wACSF) based on PAS and have used this set to predict
enthalpies of formation in the QM9 database. In order to render the training more challenging, and contrary to
Ref. \citenum{GasteggerSchwiedrzikEtAl2018JCP}, enthalpies in the dataset were neither normalised nor rescaled.
PAS were set up according to the procedure
outlined in Section \ref{sec:organic}, employing a maximum width of 8 \AA{} and using a random choice
of 10000 structures for training (and testing) as outlined in Ref. \citenum{GasteggerSchwiedrzikEtAl2018JCP},
the remainder being used as a validation set.
Mean absolute errors (MAE) over predicted enthalpies of formation for the remaining $\approx$ 124000 molecules
of the database were averaged over 5 independent partionings into train, test and validation sets as in Ref.
\citenum{GasteggerSchwiedrzikEtAl2018JCP}.

As shown in Table \ref{tbl:qm9}, while performance on the training set is comparable between the best BPSF set reported in Ref. \citenum{GasteggerSchwiedrzikEtAl2018JCP} and PAS,
the quality of the predictions on test and validation sets within a given selection of
10000 test- and training structures considerably improves when PAS are used.
For the random partitionings into training and test data studied here, errors obtained from HDNNPs trained
with PAS are about 5-fold smaller than their BPSF counterparts. It should be noted that
once BPSF are used to train a HDNNP on an appropriately normalised and rescaled training set, their performance
approaches that of PAS on a non-normalised test set\cite{GasteggerSchwiedrzikEtAl2018JCP}.
This indicates that the robustness of the predictions substantially increases when PAS are used,
as test and validation error are substantially lower even when the dataset is not subjected to any normalisation. 

\begin{table}[tbp]
\caption{MAE [kcal/mol] over training, test and validation set of the QM9 database as described in the main text.
         Note that the dataset was not normalised.}
\begin{tabular}{lrrr}
MAE  & Training & Test & Validation \\
\toprule
BPSF   & 1.51 & 11.21   & 10.72  \\
PAS    & 1.63 & 2.13    &  2.24  \\
\bottomrule
\end{tabular}
\label{tbl:qm9}
\end{table}

\subsection{Computational Footprint}

\begin{table}[bp]
	\centering
	\caption{\label{tab:speedup} MD simulation execution time of 360 water molecules (100 timesteps, 4 cores).
                                SFGroups/Caching denotes the algorithmic optimisations outlined in Ref.\
                                \citenum{SingraberBehlerEtAl2019JCTC}, values in parentheses denote speedups.}
	\begin{tabular}{lccc}
		Optimization     & BPSF          & PSF                     & PAS                     \\
		\toprule
		None             & $37.28$       & $7.72 \, (\times 4.83)$ & $8.07 \, (\times 4.62)$ \\
		SFGroups/Caching & $12.50$       & $6.82 \, (\times 1.83)$ & $6.90 \, (\times 1.81)$ \\
		\midrule
		Speedup via Opt. & $\times 2.98$ & $\times 1.13$           & $\times 1.17$           \\
		\bottomrule
	\end{tabular}
\label{tbl:performance}
\end{table}

The choice of symmetry functions has an influence on both, the computational footprint of
training as well as the cost of force predictions during production runs. For the former, if SFs can be calculated for all structures
in the training set and then stored, it is mainly the number of SFs that influences performance: More SFs imply more weights to be optimised
and as such result in higher execution times. Memory requirements will grow as well as the number of symmetry functions increases.
Even if a (generic) set of two symmetry functions might comprise the same number of input functions, pruning of functions that never exceed
a certain threshold (typically $10^{-3}$) can result in different set sizes during training. In productive MD runs, SFs have to be recalculated
for every new structure and hence, performance will not only be determined by their sheer number, but also by the number and type of floating point
operations involved in their computation. In the following, we will analyse both performance related aspects.

Table \ref{tbl:performance} lists the cost of running 100 steps of MD for a periodic box containing 360 water molecules
 using the HDNNPs investigated here both for a straightforward calculation of all SFs as well as using an optimised algorithm
tailored for distributed memory parallelisation on core processing units (CPUs) as reported in Ref.\ \onlinecite{SingraberBehlerEtAl2019JCTC}.
The former serves as a measure for the cost of the floating-point operations associated to every SF type; this corresponds to the limit
of all $N$ out of $N$ symmetry functions being independent and differently parametrised. In comparison with the former, the latter indicates 
the benefit of underlying algorithmic optimisations; this however requires that a subset of the $N$ symmetry functions contain
identical parameters and cutoff functions in order to benefit from a grouping and caching strategy.
The effective speedups obtained by using PSF or PAS for the water setup reach almost a factor of 5 when no grouping or caching strategy is applied.
This indicates that the number of expensive floating point operations is significantly reduced when purely polynomial SFs are used. \emph{Vice versa},
symmetry function grouping and caching has a much lower influence of execution times of PSF and PAS, where those optimisations reduce the computational
cost by about 15\%. As would be expected by the larger number of terms involved in the evaluation of PAS, their computation is slightly more 
expensive than the evaluation of PSF. However, the difference is small enough to be negligible in practice.

In contrast, BPSFs -- for which grouping and caching strategies were developed in the first place -- benefit from
speedups of up to a factor of 3 when grouping and caching strategies are introduced. Note that, by construction, PAS and PSF are not intended
to require use of optimisation strategies, which is reflected in comparably modest speedups upon introduction of grouping and caching.
Still, even in those cases, PSF and PAS are almost a factor of 2 faster than BPSF.
In addition, it should be noted that PAS yield the lowest RMSD on water with the smallest total number of SFs;
they therefore do not only lower the computational cost, but also the memory footprint with respect to BPSFs.

\begin{table}[tbp]
\caption{Number of symmetry functions effectively used during training of the systems investigated here
         after removing SFs with amplitudes $< 10^{-3}$. 
         BPSF-parameters were chosen according to
         Ref.\ \citenum{ImbalzanoAnelliEtAl2018JCP}.}
\begin{tabular}{lrrr}
System  & B   & PSF & PAS \\
\toprule
Water   & 57  &  62 &  54 \\
Cu$_2$S & 138 & 162 & 178 \\
DMABN   & 893 & 852 & 841 \\
Ethyl benzene & 272 & 366 & 368 \\
Anisole & 1025 & 1057 & 1036 \\
\bottomrule
\end{tabular}
\label{tbl:number}
\end{table}

Table \ref{tbl:number} compares the numbers of symmetry functions effectively used during training after pruning all SFs
that never exceed a magnitude of $10^{-3}$. This provieds a measure for computational
cost and memory requirements. Apart from ethyl benzene,
in which significantly more SFs of BPSF-type were removed during pruning, numbers are comparable
between BPSF, PSF and PAS. It should be noted that all systems except water were trained with automatically generated sets of SF
parameters, which also accounts for the different extent of functions with maximum amplitude of $< 10^{-3}$ in these setups. In this context,
it is interesting to note that in spite of the structural similarities between anisole and ethyl benzene, for BPSFs,
 there is a significant difference in the number of pruned symmetry functions. For PAS and PSF, no such difference can be observed;
this can also account for the larger errors in force RMSDs that are observed when ethyl benzene is trained with a set of BPSFs.
 Within the hand-picked set of SFs for water, differences between PSF and PAS are notable, with the 
latter allowing for a reduction of about 15\% with respect to the former. BPSFs lie in between. This suggests that, for hand-selected
sets, PAS allow for an even smaller number of SFs to be used.

Overall, the performance of PSF and BPSFs was comparable for all systems studied here. Force RMSDs for water and Cu$_2$S
were slightly higher for PSF, however, this was not reflected in the dynamic and structural properties of water obtained by
2 ns of HDNNP-MD, where results using PSF and BPSFs showed excellent agreement. On the other hand,
for DMABN and liquid ethyl benzene, PSF considerably outperformed BPSFs. For all systems, PAS were able to outperform
either PSF or BPSF. In particular, when used in the wACSF scheme\cite{GasteggerSchwiedrzikEtAl2018JCP}, PAS greatly
improve performance over BPSF for data sets that are not normalised, indicating superior robustness.
We therefore strongly recommend PAS for future HDNNP setups,
since they allow to speed up productive MD runs by a factor of 1.8 and further improve accuracy of the HDNNPs.
In particular, due to their floating point cost being lower by about a factor of almost 5,
we particularly recommend use of PAS for implementations
where caching and grouping strategies are not feasible. This can, for instance, be the case when offloading computation of
SFs to GPUs, where the resolution of \texttt{if}-statements and loop breaks associated to optimisation schemes
can considerably slow down computation with respect to purely arithmetic operations. Not least, thanks to their small computational
footprint, PAS can also be used in programs that have not undergone extensive optimisation, and their simple structure facilitates
implementation and calculation of derivatives up to an arbitrary order defined by the underlying polynomial.

\section{Conclusion and Outlook}

Here, we have introduced a new family of symmetry functions for Behler-Parrinello HDNNPs,
based on polynomials with compact support for both radial and angular environments,
which supersedes use of a separate set of radial cutoff functions.
The centres and widths of our polynomial symmetry functions can be freely chosen. This is notably the case
for angular functions, which have so far been restricted to peak at $0^\circ$ or $180^\circ$. As long as symmetry of derivatives
on $0^\circ$ and $180^\circ$ is maintained, angular functions can be centred on any point within $[0^\circ,180^\circ]$
and their width $\Delta \vartheta$ can be freely chosen. We have introduced two types of radial symmetry functions, PSF and PAS,
with the former being point-symmetric on $[r_s,r_s+\Delta]$, whereas the latter have a long-range tail reminiscent of the product
of a cutoff function and a Gaussian.
Our polynomial symmetry functions considerably simplify the choice of radial symmetry function parameters with respect to common Gaussian functions,
since the position of maxima and minima can be straightforwardly predicted without having to take into account shifts
and asymmetries introduced by a cutoff function.

The accuracy of a PSF- and PAS-based setup
 with respect to conventional BPSFs was assessed for various phases of water, for solid Cu$_2$S as well as
for organic molecules in the gaseous and liquid phase. For water, Cu$_2$ and DMABN, PSF, PAS and BPSF all showed excellent accuracy.
Structural and dynamic properties obtained from 2 ns HDNNP-MD of liquid water revealed no relevant differences between the neural network potentials
obtained with any of the three symmetry function types. In the case of off-equilibrium structures of a DMABN molecule in the gas phase,
PAS and PSF have outperformed BPSFs for force predictions by 15\% and improved energy RMSDs by 50\%.
This situation was much more prominent for liquid ethyl benzene, where force RMSDs over the test set were more than 50\% higher for 
BPSF than for PSF or PAS. Force prediction errors were generally higher for liquid anisole, however, they remained highly comparable
between all types of SFs studied here.
Generally, PSF performed on par with BPSF, except for ethyl benzene where they performed considerably better.
 For all systems studied here, PAS consistently yielded the best results.
By using PAS in weighted atom-centred SFs (wACSFs), the predictive power of a HDNNP trained on 10000 non-normalised
structures of the QM9 database was considerably improved, with mean absolute errors for training, test and validation
set being about five-fold lower than for BPSF-based wACSFs.

We have demonstrated that HDNNP-MD of liquid water is about 2 times faster using PSF or PAS compared to BPSF,
constituting a considerable improvement of performance and hence, sampling.
In terms of the floating point operations associated to the evaluation of SFs,
we found that use of PSF results in speedups of a factor of about 5 with respect to BPSF, and speedups associated
to the slightly more complex PAS still exceed a factor of 4.5. The number
of symmetry functions used for the systems investigated here were comparable between PSF, PAS and BPSF, although results
for the hand-picked set for water indicate that setups with PAS can potentially reduce the number of SFs used with respect
to common BPSFs, therefore saving on memory and on the cost of training by reducing the number of fitting parameters.
Given the substantial reduction in extensive floating point operations associated to the use of PSF and PAS, and given that accuracy
of generic sets of PAS is consistently better than that of BPSF, we advocate use of PAS as a method of choice
for future implementations. In particular, their simple structure not only simplifies their implementation and makes their setup
straightforward, but their small computational footprint can also facilitate their porting to different system architectures
without the need for extensive optimisation.

Overall, we have shown that by constructing SFs with compact support, substantial performance gains of up to a factor 
of almost 5 can be obtained without impacting
accuracy of the resulting HDNNPs. Polynomial SFs are easy to implement and set up, since they allow for a flexible choice
of radial and angular domain do not require any cutoff functions.
Our polynomial symmetry functions therefore lend themselves for all applications of HDNNPs, be it for equilibrium
or off-equilibrium stuctures of isolated molecules,
liquids or solids.

\section*{Acknowledgements}
MPB acknowledges a postdoctoral fellowship from the Swiss National Science Fondation (project 184500).
AS acknowledges support from the European Union's Horizon 2020 research and innovation programme under
Grant Agreement No.\ 676531 (project E-CAM). The results presented here were in part obtained using the Vienna Scientific Cluster (VSC).

\section*{References}
\bibliography{ms}

\end{document}


%
%
\author{Martin P. Bircher}
\affiliation{Faculty of Physics
	\\University of Vienna
	\\Boltzmanngasse 5
	\\A-1090 Wien, Austria}
\author{Andreas Singraber}
\affiliation{Institute for Theoretical Physics
	\\TU Wien - Vienna University of Technology
	\\Wiedner Hauptstra\ss e 8-10
	\\A-1040 Wien, Austria}
\affiliation{Faculty of Physics
	\\University of Vienna
	\\Boltzmanngasse 5
	\\A-1090 Wien, Austria}
\author{Christoph Dellago}
\email{christoph.dellago@univie.ac.at}
\affiliation{Faculty of Physics
	\\University of Vienna
	\\Boltzmanngasse 5
	\\A-1090 Wien, Austria}

\title{Supporting Information: Improved Description of Atomic Environments using Low-cost Polynomial Functions with Compact Support}

\maketitle

\begin{table*}
\caption{Root mean-square deviation (RMSD) of energy [meV] and force [meV/\AA{}] predictions for both, training and test
         sets, of all molecules studied here.}
\begin{tabular}{ll|c|c|c|c}
System  & SF  & RMSD($E^\text{train}_\text{nnp},E^\text{train}_\text{ref}$)
              & RMSD($E^\text{test} _\text{nnp},E^\text{test} _\text{ref}$)
              & RMSD($F^\text{train}_\text{nnp},F^\text{train}_\text{ref}$) 
              & RMSD($F^\text{test} _\text{nnp},F^\text{test} _\text{ref}$)  \\
\toprule  
                & BPSF& 1.17  & 1.17 & 43.7  & 43.7    \\
Water           & PSF & 1.14  & 1.27 & 48.0  & 48.5    \\
                & PAS & 0.95  & 1.01 & 43.3  & 43.2    \\
\midrule
                & BPSF& 1.06  & 1.14 & 51.7  & 51.9    \\
Cu$_2$S         & PSF & 1.04  & 1.15 & 58.0  & 56.7    \\
                & PAS & 0.99  & 1.01 & 52.3  & 52.3    \\
\midrule
                & BPSF& 0.41  & 0.74 & 57.6  & 63.2    \\
DMABN           & PSF & 0.21  & 0.64 & 46.9  & 56.6    \\
                & PAS & 0.19  & 0.71 & 45.0  & 54.0    \\
\midrule
                & BPSF& 0.05  & 0.22 & 87.0  & 89.5    \\
Ethyl benzene   & PSF & 0.17  & 0.25 & 67.1  & 71.2    \\
                & PAS & 0.05  & 0.22 & 58.6  & 61.7    \\
\midrule
                & BPSF& 0.08  & 0.32 & 100.3 & 100.8    \\
Anisole         & PSF & 0.11  & 0.32 & 104.9 & 106.5   \\
                & PAS & 0.11  & 0.28 & 96.7  &  98.7   \\
\bottomrule
\end{tabular}
\label{tbl:rmsd}
\end{table*}

\begin{table}
\caption{Mean absolute errors (MAE) [kcal/mol] for enthalpies of formation of the QM9 database. HDNNPs were trained
         using 5 independent partitionings of the train/test (10000 selected structures). Datasets were not normalised
         or rescaled.}
\begin{tabular}{ll|c|c|c|c|c}
SF       & Type       &  Set 1  &  Set 2  &  Set 3  &  Set 4  &  Set 5 \\ 
\toprule  
         & Training   &  1.35   &  1.52   &   1.43  &  1.47   &  1.78  \\  
BPSF     & Test       &  2.20   &  3.92   &  27.41  & 12.72   &  9.78  \\ 
         & Validation &  4.54   &  6.39   &  18.30  & 13.92   & 10.44  \\ 
\midrule
         & Training   &  1.53   &  1.64   &   1.69  &  1.60   &  1.71   \\   
PAS      & Test       &  1.97   &  2.06   &   2.46  &  2.07   &  2.08   \\ 
         & Validation &  2.03   &  2.29   &   2.33  &  2.17   &  2.36   \\ 
\bottomrule
\end{tabular}
\label{tbl:qm9}
\end{table}

\section{System Setups}

\subsection{Generation of Training Data}
All training sets that are first reported here were generated with a developmental version 
of the \texttt{CPMD}\cite{CPMD} code, using the SCAN\cite{SunRuzsinszkyEtAl2015PRL}
exchange-correlation functional and PBE\cite{PerdewBurkeErnzerhof1996PRL} pseudopotentials of the Martins-Troullier type\cite{TroullierMartins1993PRB}.
A plane wave cutoff was used throughout, and calculations on isolated systems used the Tuckerman-Martyna Poisson solver\cite{MartynaTuckerman1999JCP}
in order to decouple periodic images. A plane wave cutoff energy of 80 Ry was used throughout.
 In accordance with standard practice, hydrogen atoms in Car-Parrinello MD (CPMD) 
were replaced by deuterium, and the fictitious electronic mass was set to 600 a.u.;
the wavefunction convergence threshold was set to $10^{-7}$ a.u. on the maximum norm of the gradient.
If not stated otherwise, Nos\'e-Hoover chains\cite{MartynaKleinTuckerman1992JCP} were used to thermostat the systems.

\subsubsection{Water and Copper Sulfide}
Training sets for water were obtained from Ref.\ \onlinecite{MorawietzSingraberEtAl2016PNASUSA}.
Data for Cu$_2$S corresponds to the set of Ref.\ \onlinecite{SingraberBehlerEtAl2019JCTC}.

HDNNP-MD of 360 water molecules was performed using the LAMMPS\cite{LAMMPS} program \emph{via}
the corresponding n2p2-interface. A timestep of 1 fs was used throughout.
A pre-equilibrated snapshot of 360 liquid water molecules at experimental density
(cubic supercell with $a=b=c=22.211$ \AA$^3$) was propagated in the NVT ensemble
for a total of 2 ns at 300K, followed by 2 ns of HDNNP-MD in the NVE ensemble.

\subsubsection{DMABN}
An isolated molecule of \emph{p}--(N,N--dimethyl\-amino)\-benzo\-nitrile (DMABN) was placed in a 
15 \AA{}$^3$ orthorhombic box. 20000 steps of CPMD with a timestep of 4 a.u. were performed at 990 K.
800 structures of this trajectory were selected for the training set. In order
to obtain a non-equilibrium path along a rotation of the amino group, a harmonic restraint
with $k= 0.005$ a.u. was set up between the dihedral planes of four carbons of the phenyl ring
and a dihedral spanned by two carbons of the phenyl ring, the nitrogen as well as one carbon
of the amino group. The value of the restraint was set to grow by $0.25^\circ$ per 4 a.u.
over a total of 9950 time steps. 398 regularly spaced structures were selected along this trajectory,
with the training data containing a total of 1198 different structures.

\subsubsection{Ethyl Benzene and Anisole}
In order to obtain a diverse set of structures for these systems, Born-Oppenheimer MD 
(BOMD) simulations using a timestep of 10 a.u. were performed, with 16 molecules contained
in cubic supercells at the experimental density ($a=b=c=14.8206$ \AA$^3$ for ethyl benzene
and $a=b=c=14.24$ \AA$^3$ for anisole). BOMD of ethyl benzene was performed for 11750 steps
in the NVT ensemble at 420 K, and 468 equispaced structures of the resulting trajectory were used to constitute the
training set. BOMD of anisole was performed for 10640 steps in the NVE ensemble
at 420 K, followed by 1725 steps at 1320 K in order to sample rotations of the methoxy group.
494 equispaced structures were selected for the
training set.

\subsubsection{QM9}
Enthalpies of formation of the QM9 database were taken from Ref. \citenum{RamakrishnanEtAl2014Nature}.
Contrary to Ref.  \citenum{GasteggerSchwiedrzikEtAl2018JCP}, enthalpies were neither rescaled nor normalised.

\subsection{Training of HDNNPs}
All HDNNPs were trained with the n2p2 program package\cite{n2p2}.
Symmetry functions that do not take values above $10^{-3}$ were removed from the setups.
Threshold for force and energy updates were set to 100\% of the given RMSD on the training set,
with update candidates being randomly selected from structures where force and energy
errors meet these criteria.
The energy to force update ratio (where applicable) was 1:9. 10\% of structures in the training set
were randomly discarded at the beginning and kept for validation of the resulting
HDNNP (training \emph{vs.}\ test set). Minimisation of the weights was performed using a Kalman
filter as reported in Ref.\ \onlinecite{SingraberBehlerEtAl2019JCTC} over a total of 25 epochs
for water, Cu$_2$S, DMABN and ethyl benzen, 29 epochs for anisole and 100 epochs for the QM9 database.
Input files for training runs can be 
downloaded from the n2p2 repository: \href{https://github.com/CompPhysVienna/n2p2}{https://github.com/CompPhysVienna/n2p2}.

\section{RMSDs for Forces and Energies: Training and Test Sets}

Table \ref{tbl:rmsd} lists force and energy RMSDs obtained on training and test sets.
Data refers to the best out of 10 training runs.
Table \ref{tbl:qm9} shows training, test and validation errors for the QM9 database averaged
over 5 independent partitionings of 10000 selected structures into training and test set,
with the remainder being used as a validation set as in Ref.\ \citenum{GasteggerSchwiedrzikEtAl2018JCP}.

\begin{table}
\caption{PAS parameters for wACSF used for training the QM9 database. $7$ radial and $6\times6$ angular
         terms were used. `Set' refers to the nomenclature of the main text; type refers to the choice
         of $f_\text{P}$ within a given SF setup. The same SFs were used for all possible combinations of elements.}
\begin{tabular}{ll|c|c|c|c}
Set      & Type         &  $r_\text{min}$ [a.u.] & $r_\text{max}$ [a.u.]
                        & $\vartheta_\text{min}$ [$^\circ$] & $\vartheta_\text{max}$ [$^\circ$] \\
\toprule  
PAS      & PSF radial         &  -8    & 8      &         &        \\
         & PAS radial         &  -6    & 6      &         &        \\ 
         & PAS radial         &  -4    & 4      &         &        \\ 
         & PAS radial         &  -2    & 2      &         &        \\ 
         & PAS radial         &   0    & 4      &         &        \\ 
         & PAS radial         &   2    & 6      &         &        \\ 
         & PAS radial         &   4    & 8      &         &        \\ 
         & PSF angular narrow &  -8    & 8      &         &        \\
         & PAS angular narrow &  -6    & 6      &         &        \\ 
         & PAS angular narrow &  -4    & 4      &         &        \\ 
         & PAS angular narrow &  -2    & 2      &         &        \\ 
         & PAS angular narrow &   0    & 4      &         &        \\ 
         & PAS angular narrow &   2    & 6      &         &        \\ 
         &                  &        &        & -180.00 & 180.00  \\ 
         &                  &        &        &    0.00 & 360.00  \\ 
         &                  &        &        &    0.00 & 180.00  \\ 
         &                  &        &        &    0.00 & 120.00  \\ 
         &                  &        &        &   60.00 & 180.00  \\ 
         &                  &        &        &   49.10 & 169.10  \\ 
\bottomrule
\end{tabular}
\label{tbl:paraqm9}
\end{table}

\section{Symmetry Functions}
SF parameters for all systems studied here can be downloaded from the n2p2\cite{n2p2}
repository: \href{https://github.com/CompPhysVienna/n2p2}{https://github.com/CompPhysVienna/n2p2}.
Behler-type SF setups for water and Cu$_2$S were first reported in
Refs \onlinecite{SingraberMorawietzEtAl2019JCTC} and \onlinecite{SingraberBehlerEtAl2019JCTC}, respectively.
PAS and PSF parameters are reported in Tables \ref{tbl:paracu2s} (Cu$_2$S) and \ref{tbl:parah2o} (water).
BPSF for DMABN, anisole and ethyl benzene were obtained using the procedure of Ref. \onlinecite{ImbalzanoAnelliEtAl2018JCP},
using a cutoff of 15 a.u. Radial BPSF consists of 8 automatically generated, shifted terms per element combination
whereas wide angular functions were generated from 5 non-shifted radial terms and 8 angular terms ($\lambda = \pm 1$
and $\zeta = \{1,2,4,6\}$). Parameters for PAS and PSF are shown in Table \ref{tbl:paraorga}.
For the QM9 database, weighted atom-centred symmetry functions (wACSF) as introduced in Ref.\ \citenum{GasteggerSchwiedrzikEtAl2018JCP}
were used. The BPSF set adopted for comparison corresponds to the best set reported in Ref. \citenum{GasteggerSchwiedrzikEtAl2018JCP};
our PAS parameters are shown in Table \ref{tbl:paraqm9}.

\begin{table}
\caption{PSF and PAS parameters for DMABN, ethyl benzene and ansiole. Parameters apply to all possible combinations of elements.
         11 radial and $9 \times 6$ angular terms were included for every possible combination of neighbours.
         `Set' refers to the nomenclature of the main text; type refers to the choice
          of $f_\text{P}$ within a given SF setup.}
\begin{tabular}{ll|c|c|c|c}
Set      & Type         &  $r_\text{min}$ [a.u.] & $r_\text{max}$ [a.u.]
                        & $\vartheta_\text{min}$ [$^\circ$] & $\vartheta_\text{max}$ [$^\circ$] \\
\toprule  
PAS      & PSF radial       &  -15   & 15     &         &        \\
         & PAS radial       &  -14   & 14     &         &        \\ 
         & PAS radial       &  -12   & 12     &         &        \\ 
         & PAS radial       &  -10   & 10     &         &        \\ 
         & PAS radial       &  -8    & 8      &         &        \\ 
         & PAS radial       &  -6    & 6      &         &        \\ 
         & PAS radial       &  -5    & 5      &         &        \\ 
         & PAS radial       &  -2.5  & 7.5    &         &        \\ 
         & PAS radial       &   0    & 10     &         &        \\ 
         & PAS radial       &   2.5  & 12.5   &         &        \\ 
         & PAS radial       &   5.0  & 15.0   &         &        \\ 
         & PSF angular wide &  -15   & 15     &         &        \\
         & PAS angular wide &  -14   & 14     &         &        \\ 
         & PAS angular wide &  -12   & 12     &         &        \\ 
         & PAS angular wide &  -10   & 10     &         &        \\ 
         & PAS angular wide &  -8    & 8      &         &        \\ 
         & PAS angular wide &  -6    & 6      &         &        \\ 
         & PAS angular wide &   0    & 10     &         &        \\ 
         & PAS angular wide &   2.5  & 12.5   &         &        \\ 
         & PAS angular wide &   5.0  & 15.0   &         &        \\ 
         &                  &        &        &  -90.00 &  90.00  \\ 
         &                  &        &        &   90.00 & 270.00  \\ 
         &                  &        &        &    0.00 & 180.00  \\ 
         &                  &        &        &    0.00 & 120.00  \\ 
         &                  &        &        &   60.00 & 180.00  \\ 
         &                  &        &        &   49.10 & 169.10  \\ 
\midrule
PSF      & PSF radial       &  -15   & 15     &         &        \\
         & PSF radial       &  -13   & 13     &         &        \\ 
         & PSF radial       &  -11   & 11     &         &        \\ 
         & PSF radial       &  -9    & 9      &         &        \\ 
         & PSF radial       &  -7    & 7      &         &        \\ 
         & PSF radial       &  -5    & 5      &         &        \\ 
         & PSF radial       &  -3    & 3      &         &        \\ 
         & PSF radial       &   0    & 6      &         &        \\ 
         & PSF radial       &   3    & 9      &         &        \\ 
         & PSF radial       &   6    & 12     &         &        \\ 
         & PSF radial       &   9    & 15.0   &         &        \\ 
         & PSF angular wide &  -15   & 15     &         &        \\
         & PSF angular wide &  -13   & 13     &         &        \\
         & PSF angular wide &  -11   & 11     &         &        \\ 
         & PSF angular wide &  -9    & 9      &         &        \\ 
         & PSF angular wide &  -7    & 7      &         &        \\ 
         & PSF angular wide &  -5    & 5      &         &        \\ 
         & PSF angular wide &   0    & 10.0   &         &        \\ 
         & PSF angular wide &   2.5  & 12.5   &         &        \\ 
         & PSF angular wide &   5.0  & 15.0   &         &        \\ 
         &                  &        &        &  -90.00 &  90.00  \\ 
         &                  &        &        &   90.00 & 270.00  \\ 
         &                  &        &        &    0.00 & 180.00  \\ 
         &                  &        &        &    0.00 & 120.00  \\ 
         &                  &        &        &   60.00 & 180.00  \\ 
         &                  &        &        &   49.10 & 169.10  \\ 
\bottomrule
\end{tabular}
\label{tbl:paraorga}
\end{table}

~\\

\begin{table}
\caption{PSF and PAS parameters for copper sulfide.
         In the PAS set, 9 radial and $(3+3) \times 4$ angular terms were included per
         element combination.
         8 radial and $(3+3) \times 4$ angular functions were used in the PSF set.
         `Set' refers to the nomenclature of the main text; type refers to the choice
          of $f_\text{P}$ within a given SF setup.}
\begin{tabular}{ll|c|c|c|c}
Set      & Type         &  $r_\text{min}$ [\AA] & $r_\text{max}$ [\AA]
                        & $\vartheta_\text{min}$ [$^\circ$] & $\vartheta_\text{max}$ [$^\circ$] \\
\toprule  
PAS      & PSF radial       &  -6    & 6      &         &        \\
         & PAS radial       &  -6    & 6      &         &        \\ 
         & PAS radial       &  -5.5  & 5.5    &         &        \\ 
         & PAS radial       &  -5    & 5      &         &        \\ 
         & PAS radial       &  -4.5  & 4.5    &         &        \\ 
         & PAS radial       &  -4    & 4      &         &        \\ 
         & PAS radial       &  -2.5  & 2.5    &         &        \\ 
         & PAS radial       &  -1.25 & 3.75   &         &        \\ 
         & PAS radial       &   0    & 5      &         &        \\ 
         & PSF angular wide &  -6    & 6      &         &        \\
         & PAS angular wide &  -6    & 6      &         &        \\ 
         & PAS angular wide &  -5.5  & 5.5    &         &        \\ 
         &                  &        &        & -180.00 & 180.00  \\ 
         &                  &        &        &    0.00 & 360.00  \\ 
         &                  &        &        &   35.00 & 135.00  \\ 
         &                  &        &        &   59.10 & 159.10  \\ 
         & PSF angular narrow &  -6    & 6      &         &        \\
         & PAS angular narrow &  -6    & 6      &         &        \\ 
         & PAS angular narrow &  -5.5  & 5.5    &         &        \\ 
         &                  &        &        & -180.00 & 180.00  \\ 
         &                  &        &        &    0.00 & 360.00  \\ 
         &                  &        &        &   35.00 & 135.00  \\ 
         &                  &        &        &   59.10 & 159.10  \\ 
\midrule
PSF      & PSF radial       &  -6    & 6      &         &        \\
         & PSF radial       &  -5    & 5      &         &        \\ 
         & PSF radial       &  -4    & 4      &         &        \\ 
         & PSF radial       &  -3    & 3      &         &        \\ 
         & PSF radial       &  -2    & 2      &         &        \\ 
         & PSF radial       &  -1    & 3      &         &        \\ 
         & PSF radial       &   0    & 4      &         &        \\ 
         & PSF radial       &   1    & 5      &         &        \\ 
         & PSF angular wide &  -6    & 6      &         &        \\
         & PSF angular wide &  -4.5  & 4.5    &         &        \\ 
         & PSF angular wide &  -3.0  & 3.0    &         &        \\ 
         &                  &        &        &  -90.00 &  90.00 \\ 
         &                  &        &        &    0.00 & 180.00 \\ 
         &                  &        &        &   90.00 & 270.00 \\ 
         &                  &        &        &   59.10 & 159.10 \\ 
         & PSF angular narrow &  -6    & 6      &         &      \\
         & PSF angular narrow &  -4.5  & 4.5    &         &      \\ 
         & PSF angular narrow &  -3.0  & 3.0    &         &      \\ 
         &                  &        &        &  -90.00 &  90.00 \\ 
         &                  &        &        &    0.00 & 180.00 \\ 
         &                  &        &        &   90.00 & 270.00 \\ 
         &                  &        &        &   59.10 & 159.10 \\ 
\bottomrule
\end{tabular}
\label{tbl:paracu2s}
\end{table}

~\\

\begin{table*}
\caption{PSF and PAS parameters for water}
\begin{tabular}{ll|c|c|c|c|ll|c|c|c|c}
PSF      & Type  &  $r_\text{min}$ [\AA] & $r_\text{max}$ [\AA]
           Type  & $\vartheta_\text{min}$ [$^\circ$] & $\vartheta_\text{max}$ [$^\circ$] &
PAS      & Type  &  $r_\text{min}$ [\AA] & $r_\text{max}$ [\AA]
                        & $\vartheta_\text{min}$ [$^\circ$] & $\vartheta_\text{max}$ [$^\circ$] \\
\toprule
PSF radial          &  H H & -12.00    & 12.0  &      &           & PSF radial & H  H &-12.00   & 12.00  & & \\   
PSF radial          &  H H & -10.75    & 10.75 &      &           & PAS radial & H  H & -11.50  &  11.50 & & \\
PSF radial          &  H H &  -9.25    & 9.25  &      &           & PAS radial & H  H & -9.375  &  9.375 & & \\
PSF radial          &  H H &  -7.25    & 7.25  &      &           & PAS radial & H  H &  -6.00  &  9.60  & & \\ 
PSF radial          &  H H &   -3.4    &  7.0  &      &           & PAS radial & H  H &  -3.75  &  7.35  & & \\ 
PSF radial          &  H H &   -1.9    &  5.5  &      &           & PAS radial & H  H &  -2.25  &  5.85  & & \\ 
PSF radial          &  H H &   -0.90   &  4.50 &      &           & PAS radial & H  H &   0.00  &  4.00  & & \\ 
PSF radial          &  H H &    0.0    &  3.2  &      &           &            &      &         &        & & \\ 
PSF radial          &  O H &  -12.00   &  12.0 &      &           & PSF radial & H  O & -12.00  &  12.00 & & \\
PSF radial          &  O H &  -10.75   &  10.75&      &           & PAS radial & H  O &  -11.50 &   11.50& & \\ 
PSF radial          &  O H &   -9.25   &  9.25 &      &           & PAS radial & H  O &  -9.375 &   9.375& & \\ 
PSF radial          &  O H &   -7.25   &  7.25 &      &           & PAS radial & H  O &   -7.00 &   7.75 & & \\
PSF radial          &  O H &    -4.15  & 6.95  &      &           & PAS radial & H  O &   -4.50 &   6.30 & & \\ 
PSF radial          &  O H &    -2.7   & 4.5   &      &           & PAS radial & H  O &   -3.00 &   4.80 & & \\ 
PSF radial          &  O H &    -1.70  & 3.5   &      &           & PAS radial & H  O &   -1.70 &   3.50 & & \\ 
PSF radial          &  O H &    -0.8   & 2.6   &      &           &            &      &         &        & & \\ 
PSF radial          &  O O &    -12.0  & 12.0  &      &           & PSF radial & O  O &  -12.00 &  12.00  & & \\  
PSF radial          &  O O &    -10.75 & 10.75 &      &           & PAS radial & O  O &  -12.00 &  12.00 & & \\
PSF radial          &  O O &    -9.25  &  9.25 &      &           & PAS radial & O  O &  -11.50 &  11.70 & & \\
PSF radial          &  O O &    -7.25  &  7.25 &      &           & PAS radial & O  O &  -9.375 &   9.375& & \\
PSF radial          &  O O &     -1.0  &  8.2  &      &           & PAS radial & O  O &   -3.30 &   10.5 & & \\
PSF radial          &  O O &      0.20 &  7.30 &      &           & PAS radial & O  O &   -1.75 &   9.00 & & \\
PSF radial          &  O O &      1.3  &  6.45 &      &           & PAS radial & O  O &    2.60 &   7.75 & & \\
PSF radial          &  O O &      2.35 &  5.65 &      &           & PAS radial & O  O &    1.50 &   6.50 & & \\
PSF angular narrow & H O O & -12.0 &  12.0  &    0    &  180.0    &PAS angular narrow & H  O O &  -12.00 &  12.00 &    45.0  & 315.0\\
PSF angular narrow & H O O & -8.40 &  8.40  &   45.0  &  315.0    &PAS angular narrow & H  O O &  -12.00 &  12.00 &  -135.0  & 135.0\\
PSF angular narrow & H O O & -8.40 &  8.40  &  -135.0 &   135.0   &PAS angular narrow & H  O O &  -7.375 &  7.375 &     0.0  & 360.0\\
PSF angular narrow & H O O & -5.9  &  5.9   &    0.0  &  360.0    &PAS angular narrow & H  O O &  -7.375 &  7.375 &  -180.0  & 180.0\\
PSF angular narrow & H O O & -5.9  &  5.9   & -180.0  &  180.0    &                   &        &         &        &          &      \\
PSF angular narrow & H H O & -12.0 &  12.0  &    0    &  180.0    &PAS angular narrow & H  H O &  -12.00 &  12.00 &    45.0  & 315.0\\
PSF angular narrow & H H O & -7.25 &   7.25 &    45.0 &   315.0   &PAS angular narrow & H  H O &  -12.00 &  12.00 &  -135.0  & 135.0\\
PSF angular narrow & H H O & -7.25 &   7.25 &  -135.0 &   135.0   &PAS angular narrow & H  H O &  -10.50 &  10.50 &     0.0  & 360.0\\
PSF angular narrow & H H O & -5.9  &   5.9  &     0.0 &   360.0   &PAS angular narrow & H  H O &  -10.50 &  10.50 &  -180.0  & 180.0\\
PSF angular narrow & H H O & -5.9  &  5.9   & -180.0  &  180.0    &PAS angular narrow & H  H O &  -7.375 &  7.375 &     0.0  & 360.0\\
PSF angular narrow & H H O & -5.0  &  5.0   &    0.0  &  360.0    &PAS angular narrow & H  H O &  -7.375 &  7.375 &  -180.0  & 180.0\\
PSF angular narrow & H H O & -5.0  &  5.0   & -180.0  &  180.0    &PAS angular narrow & H  H O &  -5.25  &  5.25  &  -180.0  & 180.0\\
PSF angular narrow & H H O & -2.8  &  2.8   & -180.0  &  180.0    &                   &        &         &        &          &      \\
PSF angular narrow & O O O & -12.0 &  12.0  &    0    &  180.0    &PAS angular narrow & O  O O & -12.00  &  12.00 &    45.0  & 315.0\\
PSF angular narrow & O O O & -8.4  &  8.4   &   45.0  &  315.0    &PAS angular narrow & O  O O & -12.00  &  12.00 &  -135.0  & 135.0\\
PSF angular narrow & O O O & -8.4  &  8.4   & -135.0  &  135.0    &PAS angular narrow & O  O O &  -10.50 &  10.50 &     0.0  & 360.0\\
PSF angular narrow & O O O & -5.9  & 5.9    &    0.0  &  360.0    &PAS angular narrow & O  O O &  -10.50 &  10.50 &  -180.0  & 180.0\\
PSF angular narrow & O O O & -5.9  & 5.9    & -180.0  &  180.0    &                   &        &         &        &          &      \\
PSF angular narrow & O H O & -12.0 &  12.0  &    0    &  180.0    &PAS angular narrow & O  H O & -12.00  &  12.00 &    45.0  & 315.0\\
PSF angular narrow & O H O & -8.4  &  8.4   &   45.0  &  315.0    &PAS angular narrow & O  H O & -12.00  &  12.00 &  -135.0  & 135.0\\
PSF angular narrow & O H O & -8.4  &  8.4   & -135.0  &  135.0    &PAS angular narrow & O  H O & -10.50  &  10.50 &     0.0  & 360.0\\
PSF angular narrow & O H O & -5.9  & 5.9    &    0.0  &  360.0    &PAS angular narrow & O  H O & -10.50  &  10.50 &  -180.0  & 180.0\\
PSF angular narrow & O H O & -5.9  & 5.9    & -180.0  &  180.0    &                   &        &         &        &          &      \\
PSF angular narrow & O H H & -12.0 &  12.0  &    0    &  180.0    &PAS angular narrow & O  H H & -12.00  &  12.00 &    45.0  & 315.0\\
PSF angular narrow & O H H & -7.25 &   7.25 &    45.0 &   315.0   &PAS angular narrow & O  H H & -12.00  &  12.00 &  -135.0  & 135.0\\
PSF angular narrow & O H H & -7.25 &   7.25 &  -135.0 &   135.0   &PAS angular narrow & O  H H &  -10.50 &  10.50 &     0.0  & 360.0\\
PSF angular narrow & O H H & -5.9  & 5.9    &    0.0  &  360.0    &PAS angular narrow & O  H H &  -10.50 &  10.50 &  -180.0  & 180.0\\
PSF angular narrow & O H H & -4.35 &  4.35  &    0.0  &  360.0    &PAS angular narrow & O  H H &  -7.375 &  7.375 &     0.0  & 360.0\\
PSF angular narrow & O H H & -5.9  & 5.9    & -180.0  &  180.0    &PAS angular narrow & O  H H &  -7.375 &  7.375 &   -180.0 & 180.0 \\
PSF angular narrow & O H H & -4.35 &  4.35  &  -180.0 &  180.0    &                   &        &         &        &          &      \\
\bottomrule
\end{tabular}
\label{tbl:parah2o}
\end{table*}

\section{References}
\bibliography{supporting}